\documentclass[aps,prb,superscriptaddress,10pt,twocolumn,floatfix]{revtex4-2}
\usepackage[colorlinks=true,allcolors=blue]{hyperref}
\usepackage{amssymb}
\usepackage{amsmath}
\usepackage{float}
\usepackage{graphicx}
\usepackage[caption=false]{subfig}
\usepackage{amsfonts}
\usepackage{xcolor}
\usepackage{epsfig}
\usepackage{color}
\usepackage{bm}
\usepackage{tabularx}
\usepackage{multirow}
\usepackage{mathtools}
\usepackage{xfrac}
\usepackage{blkarray}
\usepackage{bbold}
\usepackage[mathscr]{euscript}
\usepackage{autobreak}
\usepackage{comment}
\usepackage{makecell}
\usepackage{cancel}
\usepackage{soul, color, xcolor}
\usepackage{float}
\usepackage{amsmath}

\begin{document}
	\title{Nuclear spin relaxation in solid state defect quantum bits via electron-phonon coupling in their optical excited state}
	
	\author{Gerg\H{o} Thiering}
	\email{thiering.gergo@wigner.hun-ren.hu}
	\affiliation{HUN-REN Wigner Research Centre for Physics, P.O.\ Box 49, H-1525 Budapest, Hungary}
	
	\author{Adam Gali}
	\email{gali.adam@wigner.hun-ren.hu}
	\affiliation{HUN-REN Wigner Research Centre for Physics, P.O.\ Box 49, H-1525 Budapest, Hungary}
	\affiliation{Department of Atomic Physics, Institute of Physics, Budapest University of Technology and Economics, M\H{u}egyetem rakpart 3., H-1111 Budapest, Hungary}
	\affiliation{MTA-WFK Lend\"ulet ``Momentum'' Semiconductor Nanostructures Research Group, P.O.\ Box 49, H-1525 Budapest, Hungary}
	
	\date{\today}
	
	\begin{abstract}
		Optically accessible solid state defect spins serve as a primary platform for quantum information processing, where precise control of the electron spin and ancillary nuclear spins is essential for operation. Using the nitrogen-vacancy (NV) color center in diamond as an example, we employ a combined group theory and density functional theory study to demonstrate that spin-lattice relaxation of the  $^{14}$N nuclear spin is significantly enhanced due to strong entanglement with orbital degrees of freedom in the $|^3E\rangle$ optical excited state of the defect. This mechanism is common to other solid-state defect nuclear spins with similar optical excited states. Additionally, we propose a straightforward and versatile \textit{ab initio} scheme for predicting orbital-dependent spin Hamiltonians for trigonal defects exhibiting orbital degeneracy. 
	\end{abstract}
	
	\maketitle
	
	\section{Introduction}
	\label{sec:introduction}
	
	Solid state defect quantum bits are versatile platform for quantum information processing~\cite{weber2010quantum, Wrachtrup_2006, prawer2014quantum, Zhang_2020, Zhang2020PRL, Wolfowicz2021}.
	The most studied defect quantum bit is the nitrogen-vacancy (NV) color center in diamond where the electron spin state can be initialized and readout via spin-selective non-radiative decay from its optical excited state~\cite{Doherty_2013, Gali_2019}.
	The hyperfine interaction between the electron spin and the nuclear spins of the nitrogen and proximate $^{13}$C atoms makes it possible to control the nuclear spins that can be harnessed in quantum sensing, quantum communication and quantum computation applications~\cite{Childress_2006, van_der_Sar_2012, Golter_2013, Taminiau_2014, Wrachtrup_Waldherr_2014, Lovchinsky_2016, Reiserer_2016, Abobeih_2018_1sec_Coherence, Murta_2019, Bradley_2019, Ruf_2021, VanDeStolpe2024}.
	It is a common assumption that the relaxation time of the nuclear spins is orders of magnitude longer than that of the electron spin in diamond~\cite{Dutt_2007, Fuchs_2010,  Fuchs_2011, Casanova_2016, Rosskopf_2017, Liu_2017, Soshenko_2021, Tetienne_2021, Dreau_2013} and can be employed as quantum memories.
	This assumption is based on the weak phonon-nuclear spin interaction in the electronic ground state ($|^3A_2\rangle$) of the NV center in diamond. However, NV qubits are typically read out using optical excitations to the excited state ($|^3E\rangle$). Therefore, understanding the behavior of nuclear spin qubits during NV's optical cycles requires tracking their time evolution throughout the qubit readout process.
	
	In this study, we present a theoretical framework for orbitally driven spin-lattice relaxation of nuclear spins in the NV center in diamond. We demonstrate that nuclear spin relaxation is significantly enhanced in the optical excited state due to coherent coupling with the degenerate orbital degrees of freedom. Additionally, we develop and implement a versatile \textit{ab initio} approach to compute dynamical spin-related parameters for orbital doublets in trigonal systems. The proposed mechanism is general and can be applied to other optically readable solid state defect qubits exhibiting orbital degeneracy in their excited state.
	
	
	\section{Methods}
	\label{sec:method}
	We performed \textit{ab initio} density functional theory (DFT) calculations using the \textsc{vasp} code~\cite{Kresse_1994, Kresse1996, Kresse_1996b}, as implemented by Martijn Marsman, to compute spin parameters in the optical excited state that are not directly observable. The Heyd-Scuzeria-Ernzerhof (HSE06) hybrid functional~\cite{HSE} was employed to obtain the adiabatic potential energy surface in the excited state using the $\Delta$SCF method~\cite{Gali2009}, allowing for the parameterization of the Jahn-Teller (JT) distortion. Notably, the strong electron-phonon coupling that governs the excited state, previously reported in our earlier work~\cite{Thiering2017}, is described using JT theory~\cite{Fu2009, Abtew2011, Thiering2017} and is adopted in this study. The Perdew-Burke-Ernzerhof (PBE) functional~\cite{PBE} was applied to determine the magnetic parameters, including the nuclear quadrupole tensor ($\boldsymbol{P}$)~\cite{Petrilli_1998, Ansari_2019, Swift_2020, Choudhary_2020, Auburger_2021, Pfender_2017}, the hyperfine tensor ($\boldsymbol{A}$)~\cite{Blochl_2000, Yazyev_2005, Szasz2013, Ivady_2018}, and the electronic spin-spin tensor ($\boldsymbol{D}$)~\cite{Rayson_2008, Bodrog_2013, Ivady_2018, Biktagirov_2018}, all within the framework of the projector augmented wave (PAW) method~\cite{Blochl1994} implemented in \textsc{vasp}. Standard \textsc{vasp} projectors were used for N and C species~\cite{Kresse_1999}. The calculations were carried out in a 512-atom diamond supercell with $\Gamma$-point sampling of the Brillouin zone. A plane-wave cutoff of 370~eV was applied, and atomic positions were optimized until the forces acting on ions fell below 0.01~eV/\AA.
	
	\section{Results}
	\label{sec:results}
	\subsection{Spin properties under dynamic Jahn-Teller effect} \label{ssec:DJT}
\begin{figure}[htb]
		\includegraphics[width=1\columnwidth]{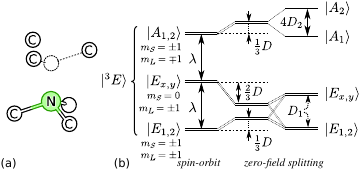}
		\caption{\label{fig:levels}%
			NV center in diamond.
			(a) Structure (b) Levels in the $^3E$ excited state for temperatures below 15~K.
			The $\lambda$, $D$, $D_2$ interactions of Eqs.~\eqref{eq:HssA} and \eqref{eq:HssE} are diagonal in the basis defined by Eq.~\eqref{eq:3EWave}, and their corresponding energy shifts are depicted. We note that $D_1$ mixes the $m_S=0$ spin projection with $m_S=\pm1$. However, this mixing is negligible, allowing us to continue using $|E_{x,y}\rangle$ and $|E_{1,2}\rangle$ as pure eigenstates, as defined in Eq.~\eqref{eq:3EWave}.}
	\end{figure}

	The NV center in diamond consists of a nitrogen atom substituting a carbon atom adjacent to a vacancy, where the defect accepts an electron from the lattice, establishing the negative charge state with a ground-state spin of $S=1$. 
	The defect has $C_{3v}$ symmetry,
	and its electronic wavefunctions are labeled using both single-group (\textit{orbital}) and double-group (\textit{orbital with spin}) notations~\cite{Doherty_2011, Maze2011}. The ground state is an orbital singlet, $|^3A_2\rangle$, whereas the optical excited state is an orbital doublet, $|^3E\rangle$. The orbital doublet forms an $E\otimes e$ JT system~\cite{Abtew2011}, meaning it exhibits strong electron-phonon coupling, which significantly affects the spin-orbit interaction~\cite{Thiering2017}.
	In adiabatic potential energy surface (APES) calculations based DFT, the JT effect manifests as a reduction 
	of the $C_{3v}$ to $C_{1h}$ symmetry, referred to as the low-
	symmetry configuration. However, the electron and phonon dynamics occur on the femtosecond timescale~\cite{Huxter_2013, Ulbricht_2016, Ulbricht_2018, Carbery_2024}, making the JT effect dynamic, so that $C_{3v}$ symmetry is effectively restored in non-ultrafast spectroscopy measurements. 
	As a result, the $|^3E\rangle$ wavefunctions are polaronically dressed states~\cite{Thiering2017, Gali_2019} but, for simplicity, we use the conventional notations: $|E_{1,2}\rangle$ ($m_S=\pm1$), $|E_{x,y}\rangle$ ($m_S=0$), $|A_1\rangle$ and $|A_2\rangle$ ($m_S=\pm1$).
	These energy levels are split due to spin-orbit interaction ($\lambda$) and various electron spin-spin interaction ($D$-tensor) parameters~\cite{PhysRevLett.102.195506, Doherty_2011, Maze2011, Maze_phd_2010}; see Fig.~\ref{fig:levels}. Notably, when strain or an external electric field is present, the degenerate $|E_{1,2}\rangle$ and $|E_{x,y}\rangle$ levels sensitively split. Additionally, in the presence of an external magnetic field, the $m_S=\pm1$ ($|E_{1,2}\rangle$) levels may undergo further splitting, ultimately determining the final energy spacing between these states~\cite{Gali_2019, Doherty_2013}. The fine structure of the $|^3A_2\rangle$ ground state is much simpler: the $m_S=0$ and $m_S=\pm1$ levels are split by magnetic spin-spin interaction, while the $m_S=\pm1$ levels are little sensitive to strain or electrical field and may experience additional splitting in the presence of a magnetic field.
	
	We express the six individual eigenstates within the $|^{3}E\rangle$ manifold as follows (see Appendices \ref{app:B} and \ref{app:C} for details)
	\begin{equation}
	\label{eq:3EWave}
	\begin{alignedat}{6}|A_{1}\rangle & = & {\textstyle \frac{1}{\sqrt{2}}}(|E_{-}^{\uparrow\uparrow}\rangle & \,-\, & |E_{+}^{\downarrow\downarrow}\rangle)\,\text{,} & \;\; & |A_{2}\rangle & = & {\textstyle \frac{1}{\sqrt{2}}}(|E_{-}^{\uparrow\uparrow}\rangle & \,+\, & |E_{+}^{\downarrow\downarrow}\rangle)\,\text{,}\\
	|E_{x}\rangle & = & {\textstyle \frac{-1}{\sqrt{2}}}(|E_{+}^{0}\rangle & \,+\, & |E_{-}^{0}\rangle)\,\text{,} &  & |E_{y}\rangle & = & {\textstyle \frac{i}{\sqrt{2}}}(|E_{+}^{0}\rangle & \,-\, & |E_{-}^{0}\rangle)\,\text{,}\\
	|E_{1}\rangle & = & {\textstyle \frac{1}{\sqrt{2}}}(|E_{-}^{\downarrow\downarrow}\rangle & \,-\, & |E_{+}^{\uparrow\uparrow}\rangle)\,\text{,} &  & |E_{2}\rangle & = & {\textstyle \frac{-i}{\sqrt{2}}}(|E_{-}^{\downarrow\downarrow}\rangle & \,+\, & |E_{+}^{\uparrow\uparrow}\rangle)\,\text{,}
	\end{alignedat}
	\end{equation}
	where the subscripts and superscripts in $|E_{m_L}^{m_S}\rangle$ denote the orbital moment and spin, respectively. At sufficiently high temperatures ($T>20$~K), the level structure of $|^3E\rangle$ 
	simplifies to that of $|^3A_2\rangle$ due to \textit{orbital averaging} activated by thermal phonons~\cite{PhysRevLett.102.195506, Goldman_PhysRevLett.114.145502, Goldman_PhysRevB.91.165201, Rogers_2009, Ulbricht_2016}. As a result, at room temperature, NV qubit protocols can address the substates of $|^3 E\rangle$ solely by their $m_S$ quantum number~\cite{Liu_2019, Toyli_2012_PhysRevX.2.031001, Gulka_2021} while the orbital degrees of freedom in Eq.~\eqref{eq:3EWave} become negligible.  This phenomenon, termed \textit{orbital averaging}, effectively causes $|^3 E\rangle$ to behave as an orbital singlet at high temperature. However, in this discussion, we retain the low-temperature level structure of $|^3E\rangle$ to explore the intricate interplay between the nuclear spins and orbitals governed by phonons.
	
	We consider the $I=1$ nuclear spin of $^{14}$N, which is more complex than the $I=1/2$ nuclear spin of $^{13}$C in the host lattice due to the presence of a quadrupole moment. The spin Hamiltonian of the ground ($g$) state in the absence of an external magnetic field is given by
	%
	\begin{equation}
	\label{eq:HssA}
	\begin{split}
	\hat{H}^{(g)}_0=&D^{(g)}\Bigl(\hat{S}_{z}^{2}-\frac{1}{3}S(S+1)\Bigr)+P^{(g)}\Bigl(\hat{I}_{z}^{2}-\frac{1}{3}I(I+1)\Bigr)\\
	&+A^{(g)}_{\parallel}\hat{S}_{z}\hat{I}_{z}+\frac{1}{2}A^{(g)}_{\perp}\left(\hat{S}_{+}\hat{I}_{-}+\hat{S}_{-}\hat{I}_{+}\right)\text{,}
	\end{split}
	\end{equation}
	where \{$\hat{S}_z$, $\hat{S}_{\pm}$\} and \{$\hat{I}_z$, $\hat{I}_{\pm}$\} represent the electron and nuclear spin operators, respectively. The spin quantization axis is chosen to be parallel to [111] symmetry axis of NV ($C_{3v}$ symmetry).
	Since $^{14}$N is positioned along the symmetry axis, the spin Hamiltonian can be fully described using only four free parameters: ($D^{(g)}, A^{(g)}_{\parallel}$, $A^{(g)}_{\perp}$, $P^{(g)}$).
	
	In the ground state, the parameters are $D^{(g)}=2.87$~GHz, $P^{(g)}=-4.96$~MHz, $A_{\parallel}^{(g)}=-2.16$~MHz, and $A_{\perp}^{(g)}=-2.70$~MHz~\cite{Gali_2008_hyperfine}, with only weakly susceptibility to strain~\cite{Udvarhelyi_2018}. The $^{14}$N nuclear spin exhibits ultralong coherence and relaxation times~\cite{Busaite_2020, Morishita_2020}, similarly to $^{13}$C nuclear spins~\cite{Maurer_2012, Scharfenberger_2014}, enabling nuclear spin readout even at room temperature~\cite{Gulka_2021}. Since $D^{(g)} \gg A^{(g)}_\perp$, nuclear spin flips are highly suppressed unless an external magnetic field compensates for the $D^{(g)}$ gap between the $m_S=\{0,-1, +1\}$ spin sublevels, a condition known as the ground state level anticrossing regime~\cite{Jaques2009_ESLAC, Gali_2009_hyperfine}. In this work, we focus on the zero or low magnetic field regime.
	
	We now analyze the optical $|^3E\rangle$ excited ($e$) state. The spin Hamiltonian for $|^3E\rangle$ consists of three terms: $\hat{H}^{(e)}=\hat{H}^{(e)}_{0}+\hat{H}^{(e)}_{\text{so}}+\hat{H}^{(e)}_{\text{orb}}$. The first term, $\hat{H}^{(e)}_{0}$, follows Eq.~\eqref{eq:HssA} with the ($g$) superscripts replaced by ($e$). The second term describes the spin-orbit interaction (see Eq.\ (6) in Ref.~\cite{Maze2011}):
	\begin{equation}
	\label{eq:HE}
	\begin{split}
	\hat{H}_{\text{so}}^{(e)}=\lambda^{(e)}\hat{\sigma}_{z}\hat{S}_{z}= & \lambda^{(e)}\Bigl(|A_{1}\rangle\langle A_{1}|+|A_{2}\rangle\langle A_{2}|\\
	& -|E_{1}\rangle\langle E_{1}|-|E_{2}\rangle\langle E_{2}|\Bigr) \text{,}
	\end{split}
	\end{equation} 
	where $\hat{\sigma}_{z}=|e_{+}\rangle\langle e_{+}|-|e_{-}\rangle\langle e_{-}|$ represent the orbital moment coupled to the electronic spin. However, in addition to these interactions, the excited state spin Hamiltonian includes orbital flipping processes described by $\hat{H}^{(e)}_{\text{orb}}$, given by
	\begin{widetext}
		\begin{equation}
		\label{eq:HssE}
		\begin{split}\hat{H}_{\text{orb}}^{(e)}= & -D_{1}^{(e)}\Bigl[\bigl(\hat{S}_{-}\hat{S}_{z}+\hat{S}_{z}\hat{S}_{-}\bigr)\hat{\sigma}_{-}+\bigl(\hat{S}_{+}\hat{S}_{z}+\hat{S}_{z}\hat{S}_{+}\bigr)\hat{\sigma}_{+}\Bigr]+D_{2}^{(e)}\Bigl[\hat{S}_{+}^{2}\hat{\sigma}_{-}+\hat{S}_{-}^{2}\hat{\sigma}_{+}\Bigr]\\
		& -P_{1}^{(e)}\Bigl[\bigl(\hat{I}_{-}\hat{I}_{z}+\hat{I}_{z}\hat{I}_{-}\bigr)\hat{\sigma}_{-}+\bigl(\hat{I}_{+}\hat{I}_{z}+\hat{I}_{z}\hat{I}_{+}\bigr)\hat{\sigma}_{+}\Bigr]+P_{2}^{(e)}\Bigl[\hat{I}_{+}^{2}\hat{\sigma}_{-}+\hat{I}_{-}^{2}\hat{\sigma}_{+}\Bigr]\\
		& -A_{1}^{(e)}\Bigl[\bigl(\hat{I}_{-}\hat{S}_{z}+\hat{I}_{z}\hat{S}_{-}+\bigr)\hat{\sigma}_{-}+\bigl(\hat{I}_{+}\hat{S}_{z}+\hat{I}_{z}\hat{S}_{+}\bigr)\hat{\sigma}_{+}\Bigr]+A_{2}^{(e)}\Bigl[\bigl(\hat{S}_{+}\hat{I}_{+}\hat{\sigma}_{-}+\hat{S}_{-}\hat{I}_{-}\hat{\sigma}_{+}\Bigr]\text{,}
		\end{split}
		\end{equation}
	\end{widetext}
	where the orbitally coupled spin-spin interaction ($\hat{\sigma}_{\pm}=|e_{\pm}\rangle\langle e_{\mp}|$) includes $D_{1,2}^{(e)}$ from Refs.~\cite{Doherty_2011, Maze2011, Maze_phd_2010}. The original Cartesian formation is converted into a representation using ladder operators ($\hat{S}_\pm = \hat{S_x} \pm i \hat{S_y}$), extending to orbitally entangled nuclear quadrupole ($P_{1,2}^{(e)}$) and hyperfine ($A_{1,2}^{(e)}$) interactions (see Appendix~\ref{app:B} for details). The parameter 
	$D_{1}^{(e)}$ couples $|E_{1,2}\rangle$ with $|E_{x,y}\rangle$, while $D_{2}^{(e)}$ splits $|A_1\rangle$ and $|A_2\rangle$ levels. In addition to the $A_\perp$ hyperfine parameter, the dynamical interactions $A_{1,2}^{(e)}$ and $P_{1,2}^{(e)}$  may also induce nuclear spin flips. The parameters $D^{(e)}=1.42$~GHz, $\lambda^{(e)}=5.3$~GHz, and $2 D_2^{(e)}=1.55$~GHz have been directly observed in photoluminescence excitation spectroscopy, while $\sqrt{2}D_1^{(e)}=200$~MHz was inferred from spin-flip optical transitions~\cite{Tamarat_2008, Doherty_2011, D1_is_SOC_wrong}.
	The hyperfine parameters $A_{\parallel}^{(e)}\approx40$~MHz and $A_{\perp}^{(e)}\approx27$~MHz for $^{14}$N have been reported~\cite{Poggiali_2017, Steiner_2010}, whereas $A_{1,2}^{(e)}$ and $P^{(e)}$, $P_1^{(e)}$ and $P_2^{(e)}$ parameters have not yet been experimentally observed, to our knowledge.
	To determine these critical parameters, we develop and implement an \textit{ab initio} theory. 
	
\begin{figure}[htb]
		\includegraphics[width=1\columnwidth]{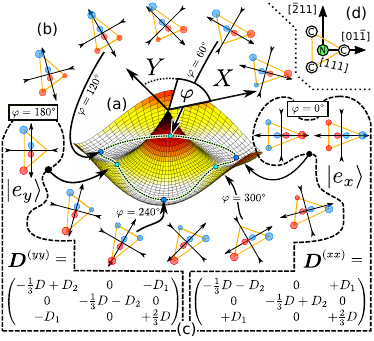}
		\caption{\label{fig:APES}%
			Dynamic Jahn-Teller effect in $|^3E\rangle$ for the NV center. (a) Adiabatic potential energy surface (APES) of the NV center in diamond, as calculated in Ref.~\cite{Thiering2017}. (b) Transformation of the occupied $|e_{\varphi}\rangle=\cos(\frac{\varphi}{2})|e_{x}\rangle-\sin(\frac{\varphi}{2})|e_{y}\rangle$ orbital, where the phase factor $e^{\pm i\varphi\pi/360^{\circ}}$ is omitted to remove the discontinuity at $\varphi=0$. The wavefunctions are shown on three carbon vacancy lobes along with the central N atom, highlighting its "$2p$" character. (c) Zero-field splitting tensors $\protect\boldsymbol{D}^{(xx)}$ and $\protect\boldsymbol{D}^{(yy)}$ at $\varphi=0^{\circ}$ and $180^{\circ}$ pseudorotations~\cite{Thompson_1985, Cocchini_1988, Bersuker_2001, bersuker2013jahn, bersuker2012vibronic}, respectively. The superscripts $^{(xx)}$, $^{(yy)}$ indicate whether $|e_x\rangle$ or $|e_y\rangle$ orbital is occupied. (d) Geometry of the undistorted system, depicting the [111] oriented coordinate system used for $\protect\overleftrightarrow{D}^{(..)}$ tensors.}
	\end{figure}

	The dynamical nature of the $|^3E\rangle$ excited state~\cite{Thiering2017Supp, bersuker2013jahn, bersuker2012vibronic} can be expressed as 
	$|e_{\varphi}\rangle=\cos(\varphi/2)|e_{x}\rangle-\sin(\varphi/2)|e_{y}\rangle$, where $\varphi$ is a (\textit{pseudo})rotation angle defined by the configurational coordinates $X$ and $Y$ on the APES; see Fig.~\ref{fig:APES}. 
	\begin{table}[] 
		\caption{\label{tab:parameters} Theoretical (experimental) spin parameters for the NV center in diamond. The labels (g) and (e) denote the ground and excited states, respectively, while "p.w." refers to the present work. The signs of the hyperfine constants are often ambiguous in experiments and are therefore not explicitly indicated here.}
		\begin{ruledtabular}
			\begin{tabular}{lrl  lrl}
				$\lambda^{(e)}            =\!\!$ &$\!\!$5.3\textsuperscript{p.w.} 4.8\footnote{\textit{Ab initio} result~\cite{Thiering2017} by means of HSE06~\cite{HSE} functional} (5.3\footnote{Expt. data vary between 5.3~GHz~\cite{PhysRevLett.102.195506} and 5.33(3)~GHz~\cite{Bassett_2014}$\!\!$}) & $\!\!\!\!\!$GHz & $\!\!\!\!\!\!\!$ $D^{(g)}            =\!\!$& 2.98\textsuperscript{p.w.}(2.87\textsuperscript{\ref{D_tensfoot}})&$\!\!\!\!\!\!\!$ GHz  
				\\ 
				\cline{1-3} 
				$A^{(g)}_\parallel            =\!\!$& $-$1.7\footnote{\label{galiHyper}\textit{Ab initio} result~\cite{Gali_2008_hyperfine} by means of LDA~\cite{Perdew_2003} (local density approximation)}(2.14\footnote{\label{EPR}Measured via electron paramagnetic resonance~\cite{Felton_2009}. Similar values were measured in Refs.~\onlinecite{Rose_2017, He_1993} for example.}) & $\!\!\!\!\!$MHz   & $\!\!\!\!\!\!\!$  $D^{(e)}            =\!\!$& 1.60\textsuperscript{p.w.}(1.42\footnote{\label{D_tensfoot}Experimental values are taken from Refs.~\onlinecite{Doherty_2011} or \onlinecite{PhysRevLett.102.195506}}) &$\!\!\!\!\!\!\!$ GHz  
				\\
				$A^{(g)}_\perp            =\!\!$& $-$1.7\textsuperscript{\ref{galiHyper}}(2.70\textsuperscript{\ref{EPR}}) & $\!\!\!\!\!\!\!$ MHz	&$\!\!\!\!\!\!\!$	$D^{(e)}_1  		=\!\!$&$\!\!$140\textsuperscript{p.w.}($200/\sqrt{2}$\textsuperscript{\ref{D_tensfoot}})& $\!\!\!\!\!$MHz  \\
				$A^{(e)}_\parallel  		=\!\!$& $+$39\textsuperscript{p.w.}(40\footnote{Measured via ESLAC, see Ref.~\onlinecite{Steiner_2010} })&$\!\!\!\!\!\!\!$ MHz 	& $\!\!\!\!\!\!\!$ $D^{(e)}_2          =\!\!$& 648\textsuperscript{p.w.}$(1550/2$\textsuperscript{\ref{D_tensfoot}})&$\!\!\!\!\!$MHz 	  
				\\
				\cline{4-6}
				$A^{(e)}_\perp          =\!\!$& $+$25\textsuperscript{p.w.}(23\footnote{Measured via ESLAC, see Ref.~\onlinecite{Poggiali_2017} })&$\!\!\!\!\!$MHz	& $\!\!\!\!\!\!\!$ $P^{(g)}            =\!\!$   &  $-$5.37\textsuperscript{\ref{Q_measurement}}($-$4.95\footnote{\label{Q_measurement}\textit{Ab initio} result~\cite{Pfender_2017} by means of HSE06~\cite{HSE} functional. Experimental result also from Ref.~\onlinecite{Pfender_2017}. Similar values were measured in Refs.~\onlinecite{Felton_2009, Rose_2017, He_1993} for example.  })& $\!\!\!\!\!$MHz \\
				$A^{(e)}_1            =\!\!$& $-$56\textsuperscript{p.w.}(n.a.) &$\!\!\!\!\!\!\!$ kHz & $\!\!\!\!\!\!\!$  $P^{(e)}            =\!\!$  &  $-$3.8\textsuperscript{p.w.}(n.a.)&$\!\!\!\!\!$MHz 
				\\
				$A^{(e)}_2            =\!\!$& $-$44\textsuperscript{p.w.}(n.a.) & $\!\!\!\!\!$kHz $\,\,$ & $\!\!\!\!\!\!\!$ $P^{(e)}_1  		=\!\!$   & +10.4\textsuperscript{p.w.}(n.a.)& $\!\!\!\!\!$kHz  
				\\
				\cline{1-3} 
				$\tau_\text{rad}^{-1}           =\!\!$&  n.a. (83\footnote{ See Refs.~\onlinecite{Robledo_2011, Toyli_2012_PhysRevX.2.031001, Goldman_PhysRevB.91.165201, Goldman_PhysRevLett.114.145502} for radiative lifetime of $|^3 E\rangle$})   & $\!\!\!\!\!$MHz & $\!\!\!\!\!\!\!$ $P^{(e)}_2          =\!\!$  & +8.2\textsuperscript{p.w.}(n.a.)& $\!\!\!\!\!$kHz  \\							
			\end{tabular} 
		\end{ruledtabular} 	
	\end{table}
	
	We focus on the second rank tensors ($\boldsymbol{\Gamma} = \{\boldsymbol{D}, \boldsymbol{A}, \boldsymbol{P}\}$), which describe spin-spin interactions.
	The occupancy of the $|e_{x}\rangle$ or $|e_{y}\rangle$ orbital is controlled by either relaxing the atomic positions to reach the JT distorted minimum or converging the system toward the saddle point, respectively. Consequently, the resulting tensors $\boldsymbol{\Gamma}^{(xx)}$ or $\boldsymbol{\Gamma}^{(yy)}$ depend on whether the $|e_{x}\rangle$ or $|e_{y}\rangle$ orbital is occupied. In principle, one can determine $\boldsymbol{\Gamma}^{(\varphi)}$ at any desired angle $\varphi$.
	However, for simplicity, we distribute a single electron equally between the $|e_x\rangle$ and $|e_y\rangle$ orbitals during the self-consistent electronic cycle of DFT calculation. We then manually enforce full occupancy of $|e_x\rangle$ while leaving $|e_y\rangle$ completely unoccupied, without further optimization of the orbitals or atomic positions, yielding the $\boldsymbol{\Gamma}^{(xx)}$ tensors as implemented in \textsc{vasp}.
	
	We find from group theory derivation (see Eq.~\eqref{eqapp:Dpars}) that diagonal terms (in the orbital degrees of freedom) appearing in Eq.~\eqref{eq:HssA} can be expressed as $\Gamma=\Gamma_{zz}^{(xx)}-(\Gamma_{xx}^{(xx)}+\Gamma_{yy}^{(xx)})/2$ for the $D$ and $P$ parameters.
	Unlike $\boldsymbol{D}$ and $\boldsymbol{P}$, the trace of $\boldsymbol{A}$ is nonzero. As a result, the hyperfine interaction is characterized by two independent parameters: $A_{\parallel}=A_{zz}^{(xx)}$ and $A_{\perp}=(A_{xx}^{(xx)}+A_{yy}^{(xx)})/2$.
	Conceptually, these parameters can be viewed as symmetrization of $\boldsymbol{\Gamma}$ tensors to the $C_{3v}$ symmetry, which also appears in the $|^3 A_2\rangle$ ground state, albeit with different values.
	
	The orbital flipping terms in Eq.~\eqref{eq:HssE} represent dynamic deviations of $\boldsymbol{\Gamma}$ tensors from $C_{3v}$ symmetry. Specifically, the presence of an $|e_{x}\rangle$ orbital spontaneously breaks the trigonal symmetry, introducing additional terms that violate $C_{3v}$ constraints. These deviations allows us
	to extract $\Gamma_{1}=\Gamma_{xz}^{(xx)}$ and $\Gamma_2=\Gamma_{xy}^{(xx)}$, respectively.
	
	However, due to the strong electron-phonon coupling, $\Gamma_1$ and $\Gamma_2$ are subject to the so-called Ham reduction~\cite{Ham_1965}, which results in partial averaging of the orbital degrees of freedom during dynamic JT motion.
	In our case, the NV center follows an $E\otimes e$ JT system, for which Bersuker provided a method to estimate the Ham reduction factors $p$ and $q$ (see Refs.~\onlinecite{bersuker2013jahn, bersuker2012vibronic} for details).
	Previously, we demonstrated that the spin-orbit coupling parameter $\lambda$ is reduced by the Ham reduction factor $p$, as $\hat{\sigma}_z$ transforms as $A_2$ under $C_{3v}$ symmetry~\cite{Thiering2017}. In our present work, we find $p=0.262$ from the HSE06 APES, see Appendix~\ref{app:E} for comparison to our previous result~\cite{Thiering2017}. 
	The interactions $\Gamma_1$ and $\Gamma_2$ transform as $E$, leading to a reduction by a factor of $q$. A known relationship between $p$ and $q$ yields $q=(1+p)/2=0.631$. This equality strictly holds only for a linear JT effect coupled to a single $E$ vibrational mode. However, it remains accurate within a few percent even when accounting for quadratic JT terms or multiple phonon mode, see Appendix~\ref{app:D} for further discussion. The parameters
	$p$ and $q$ can be interpreted as renormalization factors for physical observables~\cite{Norambuena_2020} associated with the $|^3 E\rangle$ excited state, which ultimately behaves as a quasiparticle dressed by electron-phonon interactions.
	
	The final calculated values are $P_1^{(e)}\approx10.4$~kHz, $P_2^{(e)}\approx 8.2$~kHz, $A_1^{(e)}\approx -56$~kHz, and $A_2^{(e)}\approx -44$~kHz.
	Additionally, $P^{(e)}\approx-3.8$~MHz is obtained. The values for $D^{(e)}\approx1.6$~GHz, $D_1^{(e)}\approx 140$~MHz, and $D_2^{(e)}\approx648$~MHz, as well as the hyperfine parameters $A_\parallel^{(e)}\approx39$~MHz and $A_\perp^{(e)}\approx25$~MHz, are in agreement with the known experimental data (see Table~\ref{tab:parameters} for comparison; note that the signs of these values cannot be determined experimentally).
	With all relevant parameters in hand, we proceed to compute the nuclear spin relaxation rates.
	
	\subsection{Motion of $^{14}$N nuclear spin during optical cycles}
	\label{ssec:14N}
	
\begin{figure}[htb]
		\includegraphics[width=1\columnwidth]{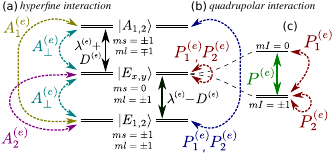}
		\caption{\label{fig:interactions}%
			Nuclear spin-flip interactions. (a) Hyperfine interactions responsible for spin transitions. (b) Quadrupolar interactions contributing to nuclear spin flips. (c) Splitting of $|E_{x,y}\rangle \otimes |m_I\rangle$ states due to the quadrupolar parameter $P^{(e)}$. The effects of $D_1$ and $D_2$ are omitted for simplicity.}
	\end{figure}

	We assume that the NV center is initialized in the $m_S=0$ spin state of the $|^3A_2\rangle$ ground state. Upon illumination, it is excited to either $|E_x\rangle$ or $|E_y\rangle$, which are often split in experiments due to strain. The electron remains in this excited for about $\tau_\mathrm{rad}\approx$~12~ns  before decaying to the ground state~\cite{Robledo_2011, Toyli_2012_PhysRevX.2.031001, Goldman_PhysRevB.91.165201, Goldman_PhysRevLett.114.145502}. During this time, we analyze the full Hamiltonian $\hat{H}^{(e)}=\hat{H}_{0}^{(e)}+\hat{H}^{(e)}_{\text{so}}+\hat{H}^{(e)}_\text{orb}$ [sum of Eqs.~\eqref{eq:HssA}, \eqref{eq:HE}, \eqref{eq:HssE}] to identify terms that may induce nuclear spin flips. The relevant terms include the hyperfine parameters $A_{\perp}^{(e)}$, $A_{1}^{(e)}$, $A_{2}^{(e)}$ and the quadrupolar parameters $P_{1}^{(e)}$, $P_{2}^{(e)}$ (see Fig.~\ref{fig:interactions}).
	
	We analyze the case where $|E_x\rangle$ is initially occupied, and the nuclear spin is in the $|0\rangle$ state. The perpendicular hyperfine interaction induces mixing via
	\begin{multline}
	\label{eq:perturbation_A_perp}
	\frac{1}{2}A_{\perp}^{(e)}\!\left(\hat{S}_{+}\hat{I}_{-}+\hat{S}_{-}\hat{I}_{+}\right) \! |E_{x}\rangle\otimes|m_{I}\!=0\rangle\\
	\begin{aligned}
	=& \frac{1}{2}A_{\perp}^{(e)}\!\Bigl(|A_{1}\rangle+|A_{2}\rangle{\textstyle +}|E_{1}\rangle-i|E_{2}\rangle\Bigr)\otimes|\!-\!1\rangle \\
	+& \frac{1}{2}A_{\perp}^{(e)}\!\Bigl(|E_{1}\rangle+i|E_{2}\rangle+|A_{1}\rangle-|A_{2}\rangle\Bigr)\otimes|\!+\!1\rangle
	\end{aligned}
	\end{multline}
	within first order perturbation. Since the radiative lifetime ($\tau_{\text{rad}}=12$~ns) of the excited state is significantly longer than the fine structure oscillation period $(\lambda^{(e)}\pm D^{(e)})^{-1}\sim2\:\text{GHz}^{-1}=0.5\:\text{ns}$ Fermi's golden rule applies; see the upcoming next section~\ref{sec:ratecalc} for further discussion. This results in a
	a nuclear spin-flip probability in the $\sim \bigl(23\text{ MHz}/2\:\text{GHz}\bigr)^{2}=1.3\times10^{-4}$ regime for $^{14}\mathrm{N}$ between \mbox{$|0\rangle$} and \mbox{$|\!\pm\! 1 \rangle$} per optical cycle. Additional hyperfine parameters, such as $A_1^{(e)}$ and $A_2^{(e)}$ are much smaller than $A_{\perp}^{(e)}$ thus their contributions are negligible. Additionally, strain parameters of individual NV centers influence the fine structure and jumping probability, leading to sample-specific variations. 
	
	Next, we consider quadrupolar interactions, specifically $P_2^{(e)}$, which affect the nuclear spin state as follows: 
	\begin{equation}
	\label{eq:perturbation_Q2}
	\begin{split}
	\!\!\!\!P_{2}^{(e)}\bigl[\hat{I}_{-}^{2}\hat{\sigma}_{-}\!+\hat{I}_{+}^{2}\hat{\sigma}_{+}\bigr]|E_{\mp}\rangle\otimes|\!\pm \! 1\rangle & = 2P_{2}|E_{\pm}\rangle\otimes|\! \mp \! 1 \rangle\\
	\!\!\!\!P_{2}^{(e)}\bigl[\hat{I}_{-}^{2}\hat{\sigma}_{-}\!+\hat{I}_{+}^{2}\hat{\sigma}_{+}\bigr]|E_{\pm}\rangle\otimes|\!\pm \! 1\rangle & = 0 \text{.}
	\end{split}
	\end{equation}
	This resembles excited-state level anticrossing (ESLAC)~\cite{Fischer_2013, Ivady_2015_PhysRevB.92.115206, Jaques2009_ESLAC, Jarmola_2020} (ESLAC) but occurs without an external magnetic field. Consequently, the perturbative approach used for Eqs.~\eqref{eq:perturbation_A_perp} is invalid, and the
	eigenstates are instead
	\begin{equation}
	\label{eq:diagonalization_Q2}
	\Psi_{\pm}=\frac{1}{\sqrt{2}}|E_{+}\rangle\otimes|\!-\!1\rangle\pm\frac{1}{\sqrt{2}}|E_{-}\rangle\otimes|\!+\!1\rangle \text{,}
	\end{equation}
	separated by $4P_{2}$ energy from each other.
	This process enables an efficient $\Delta m_I$~$=$~$\pm2$ nuclear spin-flip channel.
	However, Fermi's golden rule is violated in this case.
	In simple terms, $\tau_{\text{rad}}=12$~ns is too short to fully average out the Rabi oscillation, $\langle\sin^{2}(2\pi P_{2}^{(e)}t)\rangle=\frac{1}{2}$, with a period $1/P_2^{(e)}=58$~$\mathrm{\mu s}$, between the states $|E_{+}\rangle\otimes\!$~\mbox{$\!|\!-\!1\rangle\!$}~$\leftrightarrow |E_{-}\rangle\otimes\!$~\mbox{$\!|\!+\!1\rangle\!$}. 
	By solving the Schr\"odinger equation appropriately, we determine a small transition probability, $\int_{0}^{\infty}\frac{2\pi}{\tau_{\mathrm{rad}}}e^{-t/\tau_{\mathrm{rad}}}(2\pi 2P_{2}^{(e)}t)^{2}dt=16\pi^{2}P_{2}^{2}\tau_{\mathrm{rad}}^{2}=1.5\times10^{-6}$. 
	
	However, in typical optical qubit preparation and readout protocols, excitation into the $|^3E \rangle$ manifold is repeated multiple times, with readout times typically on the microsecond scale~\cite{Jarmola_2020, Hopper_2018, Gulka_2021, Irber_2021, Monge_2023}.
	For instance, Ref.~\onlinecite{Monge_2023} employed a 20~$\mu$s readout time while exciting the $m_S=0$ transition.
	The excitation of NV center is often saturated in this condition, meaning the system spends minimal time in the $|^3A_2\rangle$ ground state. Given $n=20$~$\mu\mathrm{s}/12$~$\text{ns}=1667$ optical cycles, the cumulative nuclear spin-flip probability mediated by $A_\perp^{(e)}$ interaction is $\sim 1667\times 1.3\times10^{-4} = 22\%$.
	This result aligns with experimental observations, where $^{15}$N nuclear spin memory degrades after approximately $1000$ consecutive optical readouts~\cite{Rosskopf_2017}.
	
	On the other hand, Rabi oscillations induced by $P_2^{(e)}$ accumulate coherently, extending the coherent evolution well beyond the excited-state lifetime $\tau_\text{rad}$. This effectively sums the total time spent in $|^3E\rangle$ over multiple optical cycles, leading to a significantly increased transition probability of $\Delta m_I = \pm 2$ nuclear spin flips, reaching $\sim40\pm20$\% (see Table~\ref{tab:fits}).
	This coherent evolution is limited only by the coherence time of $|e_{\pm}\rangle$ orbitals in the excited state. In contrast, hyperfine induced $\Delta m_I=\pm1$ transitions depend solely on the optical cycle count.
	Finally, we note that effect of $P_1^{(e)}$ is negligible in comparison to that of $A_\perp^{(e)}$ and $P_2^{(e)}$ because its impact is suppressed by $P^{(e)}$ (see Fig.~ \ref{fig:interactions}), whereas $P_{2}^{(e)}$ induces free rotation of the $^{14}$N nuclear spin. 
	
	Furthermore, the local strain parameters ($\delta_{x}$ and $\delta_{y}$) 
	may differ for each individual NV site that should be considered when our theory is compared to experimental data. Therefore, the fine structure shown in Fig.~\ref{sec:ratecalc} is further altered by the $\hat{H}_\text{strain}= -\delta_{x}\hat{\sigma}_{z}+\delta_{y}\hat{\sigma}_{x}$ interaction, see Eq.~\eqref{eq:FullHamilton} or Ref.~\onlinecite{Doherty_2013}. Due to strain, degeneracy of $|E_{x,y}\rangle$ will be lifted and also the energy positions of all $|A_{1,2}\rangle$, $|E_{1,2}\rangle$ will be slightly altered. In the readout protocol used in Ref.~\onlinecite{Monge_2023} only $|E_{x}\rangle$'s transition is being optically excited. This requires $|E_{y}\rangle$ to be spectrally distinguishable thus nonzero strain is even beneficial. However, elevated strain degrades the orbital coherence between $|E_{x,y}\rangle$, see App.~\ref{app:G} for detail. Therefore, we will derive the theory of nuclear spin transitions for an exemplary individual NV center used in Ref.~\onlinecite{Monge_2023} in the next session that exhibits strain in the GHz regime.
	
	\subsection{Theoretical nuclear spin-flip probabilities}
	\label{sec:ratecalc}
	
	\begin{figure*}
		\includegraphics[width=0.9\textwidth]{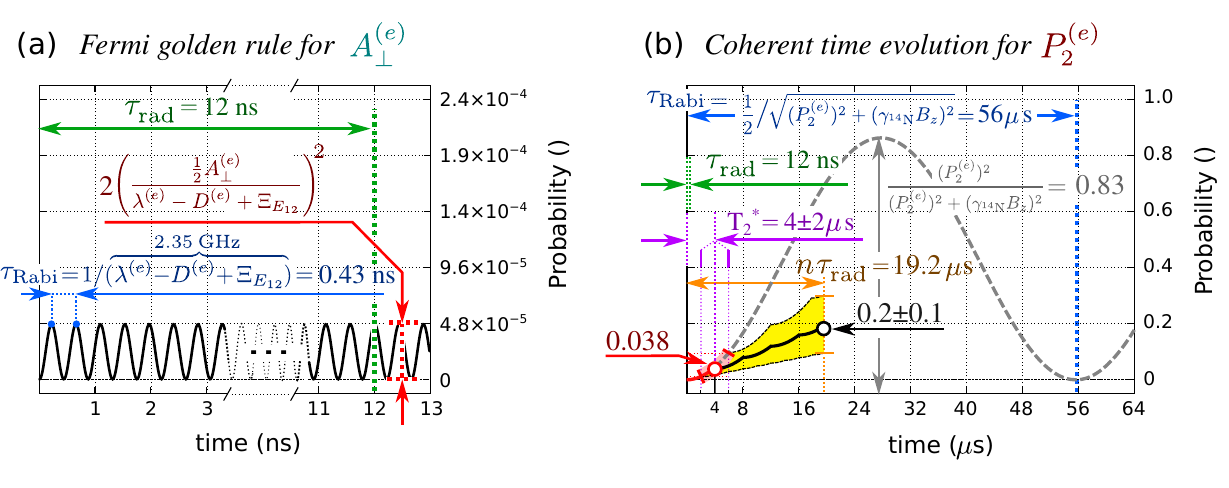}
		\caption{\label{fig:hyper}%
			Comparison of Rabi oscillations via $A_\perp^{(e)}$ hyperfine or $P_2^{(e)}$ orbital-dependent quadrupole interactions. (a) Rabi oscillations between $|E_y\rangle \otimes |m_I=0\rangle \rightarrow |E_{12}\rangle \otimes |m_I=\pm1\rangle $ quantum states driven by $A_\perp^{(e)}$. Since the radiative lifetime is not deterministic ( $\tau_\text{rad}=12$~ns), the system progressively loses coherence as the phase is randomized. Consequently, every optical cycle satisfies the conditions of Fermi's golden rule, leading to consecutive incoherent (random phase) transitions, as described in Eq.~\eqref{eq:hyperfine}. For simplicity, we assume that the 0.5~ns waiting time is sufficient for decoherence: $1/D^{(g)}=(2.88~\text{GHz})^{-1}=0.3\:\text{ns}\ll0.5\:\text{ns}=\tau_{\text{wait}}$. Thus, incoherent transitions occur twice in every optical cycle. (b) Rabi oscillations of $P_2^{(e)}$ between $|E_{y}\rangle\otimes|m_{I}=+1\rangle\rightarrow|E_{y}\rangle\otimes|-1\rangle$ quantum states,  as described in Eq.~\eqref{eq:Q1_coh}. The coherent driving is limited by the $T_2^{*}$ coherence time  (illustrated as the gray dashed curve). After each $T_2^{*} \approx \gamma_+ \approx 4\pm2$~$\mu$s interval, the phase becomes randomized due to thermal phonon-assisted decoherence (Eq.\eqref{eq:Coh1}), causing the coherent driving to restart. The yellow area represents the deviation in the final transition probability.}
	\end{figure*}

	To estimate the nuclear spin-flip probability during a saturated 20~$\mu$s readout sequence, we approximate the values based on the Supplemental Materials of Ref.~\onlinecite{Monge_2023} (e.g., Table~2). We sum the reported probabilities for different $m_S=\pm1$ states for simplicity. Experimental parameters are used where available; otherwise, we revert to theoretical values.
	
	Figure~S1(d) of Ref.~\onlinecite{Monge_2023} shows that the count rate saturates at 2$\mu$W laser power. We interpret this as the system being re-excited to the $|^3E\rangle$ manifold immediately after relaxing to $|^3A_2\rangle$, with nearly zero waiting time. Using the observed radiative lifetime of $\tau_\mathrm{rad}=12.0$~ns, we approximate the waiting time in $|^3A_2\rangle$ as $\tau_{\mathrm{wait}}=$0.5~ns. This results in approximately $n=20000/12.5=1600$ optical cycles within a single NV center over 20~$\mu$s.
	
	We first determine the nuclear spin-flip probabilities via $A_{\perp}^{(e)}$.
	We use the experimental value of strain obtained as is from Fig.~1(a,b) in Ref.~\onlinecite{Monge_2023}. Therein, the strain parameter that splits  $|E_{x}\rangle$ and $|E_{y}\rangle$ is $\delta_\perp=\frac{1}{2}\sqrt{\delta_x^2+\delta_y^2} = 4$~GHz. Thus, one may compute the exact energy positions of the full spin Hamiltonian (see Eq.~\eqref{eq:FullHamilton}) exhibiting $\delta_\perp = 4$~GHz strain. However, we used the observed values from Fig.~1(a,b) in Ref.~\onlinecite{Monge_2023} for simplicity. Therein, the energy spacing between $|E_y\rangle$, $|E_{1,2}\rangle$ levels is roughly $\approx$2.35~GHz, $|E_y\rangle$ and $|A_{1}\rangle$ is separated by $10.6$~GHz and finally the energy gap between $|E_y\rangle$, $|A_{2}\rangle$ levels is $7.6$~GHz. Utilizing the experimentally known values of $\lambda^{(e)}$, $D^{(e)}$, $D^{(e)}_2$ we determine $\Xi_{A_2}\approx2.36$~GHz, $\Xi_{A_1}\approx2.42$~GHz and $\Xi_{E_{12}}\approx-1.53$~GHz strain parameters.

	The initial state is assumed to be $m_S=0$, in $|E_x\rangle$.
	The probability of transitioning to $m_S=\pm1$ states ($|A_1\rangle$, $|A_2\rangle$, $|E_1\rangle$, and $|E_2\rangle$ multiplets) is given by: 
	\begin{equation}
	\label{eq:hyperfine}
	\begin{split}\begin{split} & p\left(|E_{x}\rangle\otimes|0\rangle\rightarrow|E_{1,2}\rangle\otimes|\pm1\rangle\right)\\
	& \qquad\quad=2\times\Biggl(\frac{\frac{1}{2}A_{\perp}^{(e)}}{\underset{2.35\:\mathrel{\text{GHz}}}{\underbrace{\lambda^{(e)}-D^{(e)}+\Xi_{E_{12}}}}}\Biggr)^{2}\times n=0.0766\text{,}
	\end{split}
	\\
	\begin{split} & p\left(|E_{x}\rangle\otimes|0\rangle\rightarrow|A_{2}\rangle\otimes|\pm1\rangle\right)\\
	& =2\times\Biggl(\frac{\frac{1}{2}A_{\perp}^{(e)}}{\underset{10.6\:\mathrel{\text{GHz}}}{\underbrace{\lambda^{(e)}+D^{(e)}+2D_{2}^{(e)}+\Xi_{A_{2}}}}}\Biggr)^{2}\times n=0.0037\text{,}
	\end{split}
	\\
	\begin{split} & p\left(|E_{x}\rangle\otimes|0\rangle\rightarrow|A_{1}\rangle\otimes|\pm1\rangle\right)\\
	& =2\times\Biggl(\frac{\frac{1}{2}A_{\perp}^{(e)}}{\underset{7.6\:\mathrel{\text{GHz}}}{\underbrace{\lambda^{(e)}+D^{(e)}-2D_{2}^{(e)}+\Xi_{A_{1}}}}}\Biggr)^{2}\times n=0.0073\text{,}
	\end{split}
	\end{split}
	\end{equation} 
	where the $2\times$ factor account for both photon absorption and luminescence events in each optical cycle, see App.~\ref{app:H} for details. 
	Thus, the total nuclear flip probability will be $2\times (2\times 0.0766+0.0018+0.0036)=0.329$ for transitions from $|0\rangle$ to $|\pm1\rangle$. We note that the inner 2$\times$ factor by the degeneracy of $|E_1\rangle$ and $|E_2\rangle$ as both states are sufficient final states to be considered. Similarly, the $|m_I=\pm1\rangle$ degeneracy causes the outer 2$\times$ factor because there are two possible final states additionally in the quantum spin. However, this 2$\times$ factor does not present in the reversed $|m_I=\pm1\rangle\rightarrow|m_I=0\rangle$ transition thus it is half as much (0.164).
	
	The nuclear spin-flip caused by quadrupolar interaction $P_2^{(e)}$
	should be calculated by solving the Schr\"odinger equation, as given by
	\begin{equation}
	\label{eq:Q1_coh}
	\begin{split}\begin{split} & p\left(|E_{y}\rangle\otimes|-1\rangle\rightarrow|E_{y}\rangle\otimes|+1\rangle\right)\\
	& =\frac{n(\tau_{\mathrm{wait}}+\tau_{\mathrm{rad}})}{T_{2}^{*}}\times\frac{(P_{2}^{(e)})^{2}}{(P_{2}^{(e)})^{2}+(\gamma_{\mathrm{^{14}N}}B_{z})^{2}}\\
	& \times\sin^{2}\left(2\pi\sqrt{(P_{2}^{(e)})^{2}+(\gamma_{\mathrm{^{14}N}}B_{z})^{2}}\frac{T_{2}^{*}\tau_{\mathrm{rad}}}{\tau_{\mathrm{wait}}+\tau_{\mathrm{rad}}}\right)
	\end{split}
	\\
	=0.191_{-0.094}^{+0.089}\approx0.2\pm0.1\text{.}
	\end{split}
	\end{equation}
	Here $T_{2}^{*}=4\pm2$~$\mu$s is an estimated coherence time, introducing an uncertainty of $\pm0.1$. The system undergoes approximately 10 coherent oscillations within $T_{2}^{*}$ before decoherence dominates. Since the applied magnetic field was 1.2~mT, the nuclear Zeeman splitting $ \gamma_{\mathrm{^{14}N}} B_z = 3.7$~kHz) is significantly smaller than $P_2^{(e)}=8.2$~kHz, having minimal influence on Rabi oscillations.
	
	Finally, we discuss the case of $P_1^{(e)}$. Firstly, one may calculate a total rate as $2\times(P_{1}^{(e)}/P^{(e)})^{2}\times n=0.0142$ similarly to that of Eq.~\eqref{eq:hyperfine}.
	However, the timescale of its Rabi oscillation is $(P^{(e)}){}^{-1}=202$~ns far exceeding $\tau_\mathrm{rad}=12$~ns thus Eq.~\eqref{eq:hyperfine} can not be used. 
	Evaluating Eq.~\eqref{eq:Q1_coh} with $P_1^{(e)}$'s case results in a negligibly small $\frac{n(\tau_{\mathrm{wait}}+\tau_{\mathrm{rad}})}{T_{2}^{*}}\times\left(\frac{P_{2}^{(e)}}{P^{(e)}}\right)^{2}\times\sin^{2}\left(2\pi P^{(e)}\frac{T_{2}^{*}\tau_{\mathrm{rad}}}{\tau_{\mathrm{wait}}+\tau_{\mathrm{rad}}}\right)\approx3\times10^{-8}$ contribution.
	
	Our calculations predict the order of magnitude of spin-flip probabilities rather than precise values which depends on external parameters such as the exact number of optical cycles. Uncertainties may also arise from symmetry constraints, choice of DFT functionals and numerical approximations (see Table~\ref{tab:functest}, where $P_{2}^{(e)}$ varies between 8.2--16.6~kHz). In addition, double $\Delta m_I=\pm1$ transitions are neglected for simplicity. We summarize our theoretical results in Table~\ref{tab:fits} where we find that our estimates are comparable to the observed nuclear spin-flip rates.
	
	\begin{table}[] 
		\caption{\label{tab:fits} Theoretical and experimental $^{14}$N spin-flip probabilities for a 20~$\mu$s readout time. The experimental data are from Table~2 in the Supplemental Materials of Ref.~\onlinecite{Monge_2023}, where we sum over the $m_S=\pm1$ spin states for clarity. Theoretical results are derived in section~\ref{sec:ratecalc}.}
		\begin{ruledtabular}
			\begin{tabular}{ccc}
				transition & theory & expt.\tabularnewline
				\hline 
				$|m_I=\pm1\rangle\rightarrow|m_I=\mp1\rangle$ & $0.2\pm0.1$ & 0.18(3)\tabularnewline
				$|m_I=\pm1\rangle\rightarrow|m_I=0\rangle$ & 0.164 & 0.16(3)\tabularnewline
				$|m_I=0\rangle\rightarrow|m_I=\pm1\rangle$ & 0.329 & 0.25(3)\tabularnewline
			\end{tabular}
		\end{ruledtabular} 	
	\end{table}

	\section{Conclusions}
	\label{sec:conclusion}
	
	In summary, (i) our \textit{ab initio} approach computes $\boldsymbol{D}$, $\boldsymbol{A}$, $\boldsymbol{P}$ tensors,  capturing non-trivial electron-phonon-mediated terms in the $|^3E\rangle$ manifold. These computed parameters can be used in Lindbladian simulations~\cite{Jin_2019, Rembold_2020, Viktor_2020, Hincks_2018, Busaite_2020, Yun_2019}. (ii) We identify significant  $\Delta m_I=\pm2$ nuclear spin-flip channels via quadrupolar term $P_{2}^{(e)}$ 
	, while hyperfine term $A_{\perp}$ 
	is responsible for $\Delta m_I=\pm1$ transitions which occurs during multiple optical cycles. (iii) Experimentally, long optical readout times ($\mu$s) may degrade qubit fidelity in $^{14}$N-based NV centers.
	Our results suggest that optical $^{14}$N nuclear spin readout should not exceed a few microseconds to minimize unwanted spin flips. 
	(iv) Finally, we note that cryogenic temperatures (4~K and lower) usually believed as an advantage for nuclear spin coherence. However, our results suggest that elevated temperatures can be beneficial to suppress decoherence processes for this specific case.

	\section*{Data Statement}
	The supporting data that covers the DFT input and output files for this article are openly available in the HUN-REN Data Repository Platform under entry~[\onlinecite{ARP_O4CEI6_2025}] where we provide a straightforward schema that can be applied on other orbitally degenerate defect systems.
	We include all DFT calculations for all $\boldsymbol{D}_\text{DFT}$, $\boldsymbol{A}_\text{DFT}$, $\boldsymbol{P}_\text{DFT}$ tensor and for all rows that of Table~\ref{tab:functest} including $\boldsymbol{R}$ rotations used in Eq.~\eqref{eq:SM:6}. 
	
	\begin{acknowledgments}
		We acknowledge Richard Monge and Carlos A.\ Meriles (CUNY–City College of New York) for discussions on experimental data and Lukas Razinkovas (FTMC, Vilnius, Lithuania) for valuable conversations on group theory.
		We gratefully acknowledge the support of the Quantum Information National Laboratory of Hungary, funded by the National Research, Development, and Innovation Office of Hungary (NKFIH) under Grant No.\ 2022-2.1.1-NL-2022-00004. G.\ T.\ was supported by the J\'anos Bolyai Research Scholarship of the Hungarian Academy of Sciences and by NKFIH under Grant No.\ STARTING 150113.
		A.\ G.\ acknowledges access to high-performance computational resources provided by KIF\"U (Governmental Agency for IT Development, Hungary) and funding from the European Commission for the QuMicro (Grant No.\ 101046911), SPINUS (Grant No.\ 101135699) and QuSPARC (Grant No.\ 101186889) projects, as well as the QuantERA II project Maestro (NKFIH Grant No.\ 2019-2.1.7-ERA-NET-2022-00045). 
	\end{acknowledgments}
	

	\appendix

	


	\section{Derivation of the spin Hamiltonian}
	\label{app:A}
	\begin{figure*}[htb]
		\includegraphics[width=\textwidth]{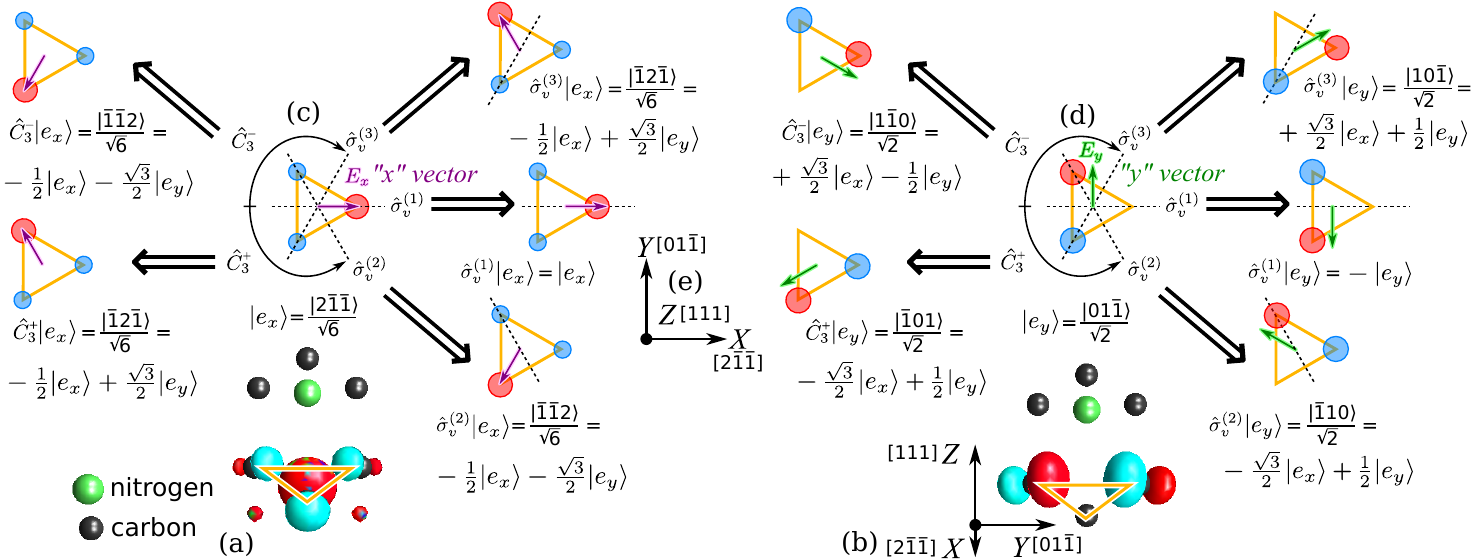}
		\caption{\label{fig:transform}%
			Transformation laws of \{$x$, $y$\} vectors, depicted as $|e_x\rangle$ (a) and $|e_x\rangle$ (b) orbitals, acting within $C_{3v}$ symmetry. The $C_3$ rotation axis points toward the [111] diamond crystal direction. (c) and (d) represent the schematic view of $|e_x \rangle$, $|e_y \rangle$ in the coordinate system depicted in (e). }
	\end{figure*}
	
	We present the six $\mathbf{M}_i$ matrices that describe the 2$\times$ degenerate \{$E_x$,$E_y$\} irreducible representation (IR) that of $C_{3v}$ point group in column IV of Eq.~\eqref{eq:transformation}. These matrices (column I in Eq.~\eqref{eq:transformation}) define the transformation that of the \{$|e_x \rangle$, $|e_y \rangle$\} hole orbitals of the NV center or the \{$x$, $y$\} vectors placed at the central point of the NV defect ( see Fig.~\ref{fig:transform} for visual interpretation). 
	The same applies to orbital operators: $\hat{\sigma}_{z}=|e_{x}\rangle\langle e_{x}|-|e_{y}\rangle\langle e_{y}|=\bigl(\begin{smallmatrix}1\\ 
	& -1
	\end{smallmatrix}\bigr)\sim"x^{2}-y^{2}"$
	and
	$\hat{\sigma}_{x}=|e_{x}\rangle\langle e_{y}|+|e_{y}\rangle\langle e_{x}|=\bigl(\begin{smallmatrix} & 1\\
	1
	\end{smallmatrix}\bigr)\sim"xy+yx"$. 
	However, we note that combination \{$-\hat{\sigma}_z$,$\hat{\sigma}_x$\} transforms similarly to \{$x$,$y$\}, as shown in column II that of Eq.~\eqref{eq:transformation}. One can determine the transformation laws of $\hat{\sigma}_i$ by applying the six symmetry operators and expanding the second order "polynomials." For example:  $\hat{\sigma}_{v}^{(1)}\bigl(-\hat{\sigma}_{z}\bigr)=\hat{\sigma}_{v}^{(1)}\bigl(|e_{y}\rangle\bigr)\hat{\sigma}_{v}^{(1)}\bigl(\langle e_{y}|\bigr)-\hat{\sigma}_{v}^{(1)}\bigl(|e_{x}\rangle\bigr)\hat{\sigma}_{v}^{(1)}\bigl(\langle e_{x}|\bigr)=+|e_{y}\rangle\langle e_{y}|-|e_{x}\rangle\langle e_{x}|=-\hat{\sigma}_{z}$.
	For completeness, we note that $\hat{\sigma}_{y}=i(|e_{y}\rangle\langle e_{x}|-|e_{x}\rangle\langle e_{y}|)=\bigl(\begin{smallmatrix} & -i\\
	i
	\end{smallmatrix}\bigr)\sim"xy-yx"$ transforms as the $A_2$ IR, while $\hat{\sigma}_{0}=(|e_{x}\rangle\langle e_{x}|+|e_{y}\rangle\langle e_{y}|)=\bigl(\begin{smallmatrix}1\\
	& 1
	\end{smallmatrix}\bigr)\sim"x^{2}+y^{2}"$ transforms as the $A_1$ IR. 
	
	\begin{widetext}
		\begin{equation}
		\begin{array}{ccc|ccc|ccc|ccc}
		& \text{I} &  &  & \text{II} &  &  & \text{III} &  &  &  & \text{IV}\\
		\hat{\mathbb{I}}\begin{pmatrix}x\\
		y
		\end{pmatrix}=\!\!\! & \mathbf{M}_{1} & \!\!\!\begin{pmatrix}x\\
		y
		\end{pmatrix} & \hat{E}\begin{pmatrix}-\hat{\sigma}_{z}\\
		+\hat{\sigma}_{x}
		\end{pmatrix}=\!\!\! & \mathbf{M}_{1} & \!\!\!\begin{pmatrix}-\hat{\sigma}_{z}\\
		+\hat{\sigma}_{x}
		\end{pmatrix} & \hat{E}\begin{pmatrix}+\hat{L}_{y}\\
		-\hat{L}_{x}
		\end{pmatrix}=\!\!\! & \mathbf{M}_{1} & \!\!\!\begin{pmatrix}+\hat{L}_{y}\\
		-\hat{L}_{x}
		\end{pmatrix} &  & \mathbf{M}_{1}= & \!\!\!\begin{pmatrix}1 & 0\\
		0 & 1
		\end{pmatrix}\\
		\hat{C}_{3}^{+}\begin{pmatrix}x\\
		y
		\end{pmatrix}=\!\!\! & \mathbf{M}_{2} & \!\!\!\begin{pmatrix}x\\
		y
		\end{pmatrix} & \hat{C}_{3}^{+}\begin{pmatrix}-\hat{\sigma}_{z}\\
		+\hat{\sigma}_{x}
		\end{pmatrix}=\!\!\! & \mathbf{M}_{2} & \!\!\!\begin{pmatrix}-\hat{\sigma}_{z}\\
		+\hat{\sigma}_{x}
		\end{pmatrix} & \hat{C}_{3}^{+}\begin{pmatrix}+\hat{L}_{y}\\
		-\hat{L}_{x}
		\end{pmatrix}=\!\!\! & \mathbf{M}_{2} & \!\!\!\begin{pmatrix}+\hat{L}_{y}\\
		-\hat{L}_{x}
		\end{pmatrix} &  & \mathbf{M}_{2}= & \!\!\!\begin{pmatrix}-\frac{1}{2} & +\frac{\sqrt{3}}{2}\\
		-\frac{\sqrt{3}}{2} & -\frac{1}{2}
		\end{pmatrix}\\
		\hat{C}_{3}^{-}\begin{pmatrix}x\\
		y
		\end{pmatrix}=\!\!\! & \mathbf{M}_{3} & \!\!\!\begin{pmatrix}x\\
		y
		\end{pmatrix} & \hat{C}_{3}^{-}\begin{pmatrix}-\hat{\sigma}_{z}\\
		+\hat{\sigma}_{x}
		\end{pmatrix}=\!\!\! & \mathbf{M}_{3} & \!\!\!\begin{pmatrix}-\hat{\sigma}_{z}\\
		+\hat{\sigma}_{x}
		\end{pmatrix} & \hat{C}_{3}^{-}\begin{pmatrix}+\hat{L}_{y}\\
		-\hat{L}_{x}
		\end{pmatrix}=\!\!\! & \mathbf{M}_{3} & \!\!\!\begin{pmatrix}+\hat{L}_{y}\\
		-\hat{L}_{x}
		\end{pmatrix} &  & \mathbf{M}_{3}= & \!\!\!\begin{pmatrix}-\frac{1}{2} & -\frac{\sqrt{3}}{2}\\
		+\frac{\sqrt{3}}{2} & -\frac{1}{2}
		\end{pmatrix}\\
		\hat{\sigma}_{v}^{(1)}\begin{pmatrix}x\\
		y
		\end{pmatrix}=\!\!\! & \mathbf{M}_{4} & \!\!\!\begin{pmatrix}x\\
		y
		\end{pmatrix} & \hat{\sigma}_{v}^{(1)}\begin{pmatrix}-\hat{\sigma}_{z}\\
		+\hat{\sigma}_{x}
		\end{pmatrix}=\!\!\! & \mathbf{M}_{4} & \!\!\!\begin{pmatrix}-\hat{\sigma}_{z}\\
		+\hat{\sigma}_{x}
		\end{pmatrix} & \hat{\sigma}_{v}^{(1)}\begin{pmatrix}+\hat{L}_{y}\\
		-\hat{L}_{x}
		\end{pmatrix}=\!\!\! & \mathbf{M}_{4} & \!\!\!\begin{pmatrix}+\hat{L}_{y}\\
		-\hat{L}_{x}
		\end{pmatrix} &  & \mathbf{M}_{4}= & \!\!\!\begin{pmatrix}1 & 0\\
		0 & -1
		\end{pmatrix}\\
		\hat{\sigma}_{v}^{(2)}\begin{pmatrix}x\\
		y
		\end{pmatrix}=\!\!\! & \mathbf{M}_{5} & \!\!\!\begin{pmatrix}x\\
		y
		\end{pmatrix} & \hat{\sigma}_{v}^{(2)}\begin{pmatrix}-\hat{\sigma}_{z}\\
		+\hat{\sigma}_{x}
		\end{pmatrix}=\!\!\! & \mathbf{M}_{5} & \!\!\!\begin{pmatrix}-\hat{\sigma}_{z}\\
		+\hat{\sigma}_{x}
		\end{pmatrix} & \hat{\sigma}_{v}^{(2)}\begin{pmatrix}+\hat{L}_{y}\\
		-\hat{L}_{x}
		\end{pmatrix}=\!\!\! & \mathbf{M}_{5} & \!\!\!\begin{pmatrix}+\hat{L}_{y}\\
		-\hat{L}_{x}
		\end{pmatrix} &  & \mathbf{M}_{5}= & \!\!\!\begin{pmatrix}-\frac{1}{2} & -\frac{\sqrt{3}}{2}\\
		-\frac{\sqrt{3}}{2} & +\frac{1}{2}
		\end{pmatrix}\\
		\hat{\sigma}_{v}^{(3)}\begin{pmatrix}x\\
		y
		\end{pmatrix}=\!\!\! & \mathbf{M}_{6} & \!\!\!\begin{pmatrix}x\\
		y
		\end{pmatrix} & \hat{\sigma}_{v}^{(3)}\begin{pmatrix}-\hat{\sigma}_{z}\\
		+\hat{\sigma}_{x}
		\end{pmatrix}=\!\!\! & \mathbf{M}_{6} & \!\!\!\begin{pmatrix}-\hat{\sigma}_{z}\\
		+\hat{\sigma}_{x}
		\end{pmatrix} & \hat{\sigma}_{v}^{(3)}\begin{pmatrix}+\hat{L}_{y}\\
		-\hat{L}_{x}
		\end{pmatrix}=\!\!\! & \mathbf{M}_{6} & \!\!\!\begin{pmatrix}+\hat{L}_{y}\\
		-\hat{L}_{x}
		\end{pmatrix} &  & \mathbf{M}_{6}= & \!\!\!\begin{pmatrix}-\frac{1}{2} & +\frac{\sqrt{3}}{2}\\
		+\frac{\sqrt{3}}{2} & +\frac{1}{2}
		\end{pmatrix}
		\end{array}
		\label{eq:transformation}
		\end{equation}
	\end{widetext}
	
	Next, we determine the transformation laws for the spin operators.
	For simplicity, we consider that the orbital angular momentum operators $\hat{L}_{x},\hat{L}_{y},\hat{L}_{z}$ transform in the same way as the spin operators $\hat{S}_{x},\hat{S}_{y},\hat{S}_{z}$. Thus, we first derive the transformation properties of $\hat{L}_{i}$ and then apply these properties to $\hat{S}_{i}$. The definitions of $\hat{L}_{i}$ are as follows:
	$\frac{i}{\hbar}\hat{L}_{x}=+y\partial_{z}-z\partial_{y}$,
	$\frac{i}{\hbar}\hat{L}_{y}=+z\partial_{x}-x\partial_{z}$,
	$\frac{i}{\hbar}\hat{L}_{z}=+x\partial_{y}-y\partial_{x}$.
	We note that the coordinate operator $z$ and the derivative $\partial_z$ both transform as the $A_1$ irreducible representation (IR), while {$\partial_x$, $\partial_y$} transform in the same way as the degenerate $E$ IR of {$x$, $y$}.
	
	Next, the transformation laws can be obtained in a manner similar to that of $\hat{\sigma}_i$. For example:
	$\hat{\sigma}_{v}^{(1)}(x\partial_{y}-y\partial_{x})=\hat{\sigma}_{v}^{(1)}(x)\hat{\sigma}_{v}^{(1)}(\partial_{y})-\hat{\sigma}_{v}^{(1)}(y)\hat{\sigma}_{v}^{(1)}(\partial_{x})=-(x\partial_{y}-y\partial_{x})$.
	The operator $\hat{S}_{z}$ transforms as the $A_{2}$ IR because $\hat{\sigma}_{v}^{(1)}\big(\hat{L}_{z}\big)=\hat{\sigma}_{v}^{(1)}(x\partial_{y}-y\partial_{x})=(-x\partial_{y}+y\partial_{x})=-\hat{L}_{z}$.
	However, one may observe that the order of $\hat{L}_{x}$ and $\hat{L}_{y}$ is not the same as that of $x$ and $y$.
	To put it simply, the transformation behavior of $\frac{i}{\hbar}\hat{L}_{x} = +y\partial_{z} - z\underline{\partial_{y}}$ is similar to that of $+y$, while $\frac{i}{\hbar}\hat{L}_{y} = +z\underline{\partial_{x}} - x\partial_{z}$ transforms as $-x$, where we use the underlining ($\underline{\partial_i}$) to emphasize the key components. Therefore, one arrives at the conclusion that the combination {$+\hat{L}_y$, $+\hat{L}_x$} transforms as {$x$, $y$}, as shown in column III that of Eq.~\eqref{eq:transformation}.
	
	Next, we will use our knowledge to determine the possible second-order polynomials involving spin operators. As an example, we derive the combination of terms that includes both $\hat{S}_z$ and $\hat{S}_{x,y}$ at first order. To facilitate this, we introduce the following $\mathbf{U}$ basis transformation matrix:
	$\begin{pmatrix}\hat{S}_{x}\\
	\hat{S}_{y}
	\end{pmatrix}=\begin{pmatrix}0 & +1\\
	-1 & 0
	\end{pmatrix}\begin{pmatrix}-\hat{S}_{y}\\
	+\hat{S}_{x}
	\end{pmatrix}=\mathbf{U}\begin{pmatrix}-\hat{S}_{y}\\
	+\hat{S}_{x}
	\end{pmatrix}$ 
	This transformation allows us to utilize the transformation laws described in column III that of Eq.~\eqref{eq:transformation}. Consequently, we find that the pair {$\hat{S}_x \hat{S}_z$, $\hat{S}_y \hat{S}_z$} transforms in the same way as {$x$, $y$}:
	\begin{widetext}
		\begin{equation}
		\begin{array}{cccc}
		\hat{E}\!\begin{pmatrix}\hat{S}_{x}\hat{S}_{z}\\
		\hat{S}_{y}\hat{S}_{z}
		\end{pmatrix}\!=\!\!\!\! & \mathbf{U}\mathbf{M}_{2}\mathbf{U}^{-1}\!\!\begin{pmatrix}\hat{S}_{x}\\
		\hat{S}_{y}
		\end{pmatrix}\underset{+\hat{S}_{z}}{\underbrace{\hat{E}(\hat{S}_{z})}}=\!\!\!\!\! & \begin{pmatrix}0 & +1\\
		-1 & 0
		\end{pmatrix}\begin{pmatrix}1 & 0\\
		0 & 1
		\end{pmatrix}\begin{pmatrix}0 & -1\\
		+1 & 0
		\end{pmatrix}(+1)\begin{pmatrix}\hat{S}_{x}\hat{S}_{z}\\
		\hat{S}_{y}\hat{S}_{z}
		\end{pmatrix} & \!\!\!\!\!=\mathbf{M}_{1}\begin{pmatrix}\hat{S}_{x}\hat{S}_{z}\\
		\hat{S}_{y}\hat{S}_{z}
		\end{pmatrix}\\
		\hat{C}_{3}^{+}\!\begin{pmatrix}\hat{S}_{x}\hat{S}_{z}\\
		\hat{S}_{y}\hat{S}_{z}
		\end{pmatrix}\!=\!\!\! & \mathbf{U}\mathbf{M}_{2}\mathbf{U}^{-1}\!\!\begin{pmatrix}\hat{S}_{x}\\
		\hat{S}_{y}
		\end{pmatrix}\underset{+\hat{S}_{z}}{\underbrace{\hat{C}_{3}^{+}(\hat{S}_{z})}}=\!\!\!\!\! & \begin{pmatrix}0 & +1\\
		-1 & 0
		\end{pmatrix}\begin{pmatrix}-\frac{1}{2} & +\frac{\sqrt{3}}{2}\\
		-\frac{\sqrt{3}}{2} & -\frac{1}{2}
		\end{pmatrix}\begin{pmatrix}0 & -1\\
		+1 & 0
		\end{pmatrix}(+1)\begin{pmatrix}\hat{S}_{x}\hat{S}_{z}\\
		\hat{S}_{y}\hat{S}_{z}
		\end{pmatrix} & \!\!\!\!\!=\mathbf{M}_{2}\begin{pmatrix}\hat{S}_{x}\hat{S}_{z}\\
		\hat{S}_{y}\hat{S}_{z}
		\end{pmatrix}\\
		\hat{C}_{3}^{-}\!\begin{pmatrix}\hat{S}_{x}\hat{S}_{z}\\
		\hat{S}_{y}\hat{S}_{z}
		\end{pmatrix}\!=\!\!\!\! & \mathbf{U}\mathbf{M}_{3}\mathbf{U}^{-1}\!\!\begin{pmatrix}\hat{S}_{x}\\
		\hat{S}_{y}
		\end{pmatrix}\underset{+\hat{S}_{z}}{\underbrace{\hat{C}_{3}^{-}(\hat{S}_{z})}}=\!\!\!\!\! & \begin{pmatrix}0 & +1\\
		-1 & 0
		\end{pmatrix}\begin{pmatrix}-\frac{1}{2} & -\frac{\sqrt{3}}{2}\\
		+\frac{\sqrt{3}}{2} & -\frac{1}{2}
		\end{pmatrix}\begin{pmatrix}0 & -1\\
		+1 & 0
		\end{pmatrix}(+1)\begin{pmatrix}\hat{S}_{x}\hat{S}_{z}\\
		\hat{S}_{y}\hat{S}_{z}
		\end{pmatrix} & \!\!\!\!\!=\mathbf{M}_{3}\begin{pmatrix}\hat{S}_{x}\hat{S}_{z}\\
		\hat{S}_{y}\hat{S}_{z}
		\end{pmatrix}\\
		\hat{\sigma}_{v}^{(1)}\!\begin{pmatrix}\hat{S}_{x}\hat{S}_{z}\\
		\hat{S}_{y}\hat{S}_{z}
		\end{pmatrix}\!=\!\!\!\! & \mathbf{U}\mathbf{M}_{4}\mathbf{U}^{-1}\!\!\begin{pmatrix}+\hat{S}_{x}\\
		-\hat{S}_{y}
		\end{pmatrix}\underset{-\hat{S}_{z}}{\underbrace{\hat{\sigma}_{v}^{(1)}(\hat{S}_{z})}}=\!\!\!\!\! & \begin{pmatrix}0 & +1\\
		-1 & 0
		\end{pmatrix}\begin{pmatrix}1 & 0\\
		0 & -1
		\end{pmatrix}\begin{pmatrix}0 & -1\\
		+1 & 0
		\end{pmatrix}(-1)\begin{pmatrix}\hat{S}_{x}\hat{S}_{z}\\
		\hat{S}_{y}\hat{S}_{z}
		\end{pmatrix} & \!\!\!\!\!=\mathbf{M}_{4}\begin{pmatrix}\hat{S}_{x}\hat{S}_{z}\\
		\hat{S}_{y}\hat{S}_{z}
		\end{pmatrix}\\
		\hat{\sigma}_{v}^{(2)}\!\begin{pmatrix}\hat{S}_{x}\hat{S}_{z}\\
		\hat{S}_{y}\hat{S}_{z}
		\end{pmatrix}\!=\!\!\!\! & \mathbf{U}\mathbf{M}_{5}\mathbf{U}^{-1}\!\!\begin{pmatrix}\hat{S}_{x}\\
		\hat{S}_{y}
		\end{pmatrix}\underset{-\hat{S}_{z}}{\underbrace{\hat{\sigma}_{v}^{(2)}(\hat{S}_{z})}}=\!\!\!\!\! & \begin{pmatrix}0 & +1\\
		-1 & 0
		\end{pmatrix}\begin{pmatrix}-\frac{1}{2} & -\frac{\sqrt{3}}{2}\\
		-\frac{\sqrt{3}}{2} & +\frac{1}{2}
		\end{pmatrix}\begin{pmatrix}0 & -1\\
		+1 & 0
		\end{pmatrix}(-1)\begin{pmatrix}\hat{S}_{x}\hat{S}_{z}\\
		\hat{S}_{y}\hat{S}_{z}
		\end{pmatrix} & \!\!\!\!\!=\mathbf{M}_{5}\begin{pmatrix}\hat{S}_{x}\hat{S}_{z}\\
		\hat{S}_{y}\hat{S}_{z}
		\end{pmatrix}\\
		\hat{\sigma}_{v}^{(3)}\!\begin{pmatrix}\hat{S}_{x}\hat{S}_{z}\\
		\hat{S}_{y}\hat{S}_{z}
		\end{pmatrix}\!=\!\!\!\! & \mathbf{U}\mathbf{M}_{6}\mathbf{U}^{-1}\!\!\begin{pmatrix}\hat{S}_{x}\\
		\hat{S}_{y}
		\end{pmatrix}\underset{-\hat{S}_{z}}{\underbrace{\hat{\sigma}_{v}^{(3)}(\hat{S}_{z})}}=\!\!\!\!\! & \begin{pmatrix}0 & +1\\
		-1 & 0
		\end{pmatrix}\begin{pmatrix}-\frac{1}{2} & +\frac{\sqrt{3}}{2}\\
		+\frac{\sqrt{3}}{2} & +\frac{1}{2}
		\end{pmatrix}\begin{pmatrix}0 & -1\\
		+1 & 0
		\end{pmatrix}(-1)\begin{pmatrix}\hat{S}_{x}\hat{S}_{z}\\
		\hat{S}_{y}\hat{S}_{z}
		\end{pmatrix} & \!\!\!\!\!=\mathbf{M}_{6}\begin{pmatrix}\hat{S}_{x}\hat{S}_{z}\\
		\hat{S}_{y}\hat{S}_{z}
		\end{pmatrix}
		\end{array}
		\label{eq:SzSxy}
		\end{equation}
	\end{widetext}
	Therefore, the combination $(\hat{S}_{x}\hat{S}_{z}+\hat{S}_{z}\hat{S}_{x})$ transforms similarly to $E_{x}$, while $(\hat{S}_{y}\hat{S}_{z}+\hat{S}_{z}\hat{S}_{y})$ transforms similarly
	to $E_{y}$. In the same way, we could determine the second-order terms involving \{$+\hat{S}_y$,$+\hat{S}_x$\} as before. However, as an example, we take an alternative approach by projecting possible combinations of $\hat{S}_x$ and $\hat{S}_y$ onto the irreducible representations $\Gamma=\{A_1,A_2,E\}$ of the $C_{3v}$ point group of the defect. The generalized projection operator $P_{ij}^{(\Gamma)}$ is defined as follows (see Eq.~(4.52) in Ref.~\onlinecite{Elliott_1979}):
	\begin{equation}
	P_{ij}^{(\Gamma)}=\frac{l_{\Gamma}}{h}\sum_{n=1}^h (M_{n}^{(\Gamma)})_{ij}\hat{O}_{n} \text{,}
	\label{eq:Projection}
	\end{equation}
	where $\hat{O}_{n}$ is the matrix representation of the six symmetry operations $\hat{O}_n=\{\hat{E}$, $\hat{C}_{3}^{+}$, $\hat{C}_{3}^{-}$, $\hat{\sigma}_{v}^{(1)}$, $\hat{\sigma}_{v}^{(2)}$, $\hat{\sigma}_{v}^{(3)} \}$, acting on the four dimensional direct product of spin operators: $\{\hat{S}_{x}$, $\hat{S}_{y}\}\otimes\{\hat{S}_{x}$, $\hat{S}_{y}\}=\{\hat{S}_{x}\hat{S}_{x}$, $\hat{S}_{x}\hat{S}_{y}$, $\hat{S}_{y}\hat{S}_{x}$, $\hat{S}_{y}\hat{S}_{y}\}$
	\begin{widetext}
		\begin{equation}
		\begin{array}{ccc}
		\hat{O}_{1}=\begin{pmatrix}+1\\
		& +1\\
		&  & +1\\
		&  &  & +1
		\end{pmatrix} & \qquad\hat{O}_{2}=\frac{1}{4}\begin{pmatrix}+1 & -\sqrt{3} & -\sqrt{3} & +3\\
		+\sqrt{3} & +1 & -3 & -\sqrt{3}\\
		+\sqrt{3} & -3 & +1 & -\sqrt{3}\\
		+3 & -\sqrt{3} & -\sqrt{3} & +1
		\end{pmatrix} & \qquad\hat{O}_{3}=\frac{1}{4}\begin{pmatrix}+1 & +\sqrt{3} & +\sqrt{3} & +3\\
		-\sqrt{3} & +1 & -3 & +\sqrt{3}\\
		-\sqrt{3} & -3 & +1 & +\sqrt{3}\\
		+3 & -\sqrt{3} & -\sqrt{3} & +1
		\end{pmatrix}\\
		\hat{O}_{4}=\begin{pmatrix}+1\\
		& -1\\
		&  & -1\\
		&  &  & +1
		\end{pmatrix} & \qquad\hat{O}_{5}=\frac{1}{4}\begin{pmatrix}+1 & +\sqrt{3} & +\sqrt{3} & +3\\
		+\sqrt{3} & -1 & +3 & -\sqrt{3}\\
		+\sqrt{3} & +3 & -1 & -\sqrt{3}\\
		+3 & -\sqrt{3} & -\sqrt{3} & +1
		\end{pmatrix} & \qquad\hat{O}_{6}=\frac{1}{4}\begin{pmatrix}+1 & -\sqrt{3} & -\sqrt{3} & +3\\
		-\sqrt{3} & -1 & +3 & +\sqrt{3}\\
		-\sqrt{3} & +3 & -1 & +\sqrt{3}\\
		+3 & +\sqrt{3} & +\sqrt{3} & +1
		\end{pmatrix}
		\end{array}
		\end{equation}
		where we take advantage of the fact that $\{\hat{S}_{x}$, $\hat{S}_{y}\}$ transforms as $(\mathbf{U}\mathbf{M}_{n}\mathbf{U}^{-1})$. Thus, the Kronecker product we seek will transform as $\hat{O}_{n}=(\mathbf{U}\mathbf{M}_{n}\mathbf{U}^{-1})\otimes(\mathbf{U}\mathbf{M}_{n}\mathbf{U}^{-1})$. Therefore, the symmetrized bases are as follows:
		\begin{equation}
		\begin{aligned}
		\label{eq:SxSy}
		|A_{1}\rangle\langle A_{1}|=P_{11}^{(A_{1})}=\frac{1}{6}\!\left[\sum_{n}^{6}\chi_{n}^{(A_{1})}\hat{O}_{n}\right]= & \frac{1}{2}\!\begin{pmatrix}+1 &  &  & +1\\
		& 0 & 0\\
		& 0 & 0\\
		+1 &  &  & +1
		\end{pmatrix} &  & \!\!\!\!\!\Rightarrow|A_{1}\rangle\langle A_{1}|\begin{pmatrix}a\\
		b\\
		c\\
		d
		\end{pmatrix}=\frac{a+d}{2}\!\begin{pmatrix}+1\\
		0\\
		0\\
		+1
		\end{pmatrix}\Longrightarrow\!\!\!\!\! & \boxed
		{
			\begin{aligned}
			& "A_{1}"=\\
			& {\textstyle \frac{1}{2}}(\hat{S}_{x}^{2}+\hat{S}_{y}^{2})
			\end{aligned}
		}
		\\
		|A_{2}\rangle\langle A_{2}|=P_{11}^{(A_{2})}=\frac{1}{6}\!\left[\sum_{n}^{6}\chi_{n}^{(A_{2})}\hat{O}_{n}\right]= & \frac{1}{2}\!\begin{pmatrix}0\\
		& +1 & -1\\
		& -1 & +1\\
		&  &  & 0
		\end{pmatrix} &  & \!\!\!\!\!\Rightarrow|A_{2}\rangle\langle A_{2}|\begin{pmatrix}a\\
		b\\
		c\\
		d
		\end{pmatrix}=\frac{b-c}{2}\!\begin{pmatrix}0\\
		+1\\
		-1\\
		0
		\end{pmatrix}\Longrightarrow\!\!\!\!\! & \boxed{\begin{aligned} & "A_{2}"=\\
			& {\textstyle \frac{1}{2}}(\hat{S}_{x}\hat{S}_{y}-\hat{S}_{y}\hat{S}_{x})
			\end{aligned}
		}\\
		|E_{x}\rangle\langle E_{x}|=P_{11}^{(E)}=\frac{2}{6}\!\left[\sum_{n}^{6}(\mathbf{M}_{n})_{11}\hat{O}_{n}\right]= & \frac{1}{2}\!\begin{pmatrix}-1 &  &  & +1\\
		& 0 & 0\\
		& 0 & 0\\
		+1 &  &  & -1
		\end{pmatrix} &  & \!\!\!\!\!\Rightarrow|E_{x}\rangle\langle E_{x}|\begin{pmatrix}a\\
		b\\
		c\\
		d
		\end{pmatrix}=\frac{a-d}{2}\!\begin{pmatrix}-1\\
		0\\
		0\\
		+1
		\end{pmatrix}\Longrightarrow\!\!\!\!\! & \boxed{\begin{aligned} & "E_{x}"=\\
			& {\textstyle \frac{1}{2}}(-\hat{S}_{x}^{2}+\hat{S}_{y}^{2})
			\end{aligned}
		}\\
		|E_{y}\rangle\langle E_{x}|=P_{21}^{(E)}=\frac{2}{6}\!\left[\sum_{n}^{6}(\mathbf{M}_{n})_{21}\hat{O}_{n}\right]= & \frac{1}{2}\!\begin{pmatrix}0 &  &  & 0\\
		-1 & 0 & 0 & +1\\
		-1 & 0 & 0 & +1\\
		0 &  &  & 0
		\end{pmatrix} &  & \!\!\!\!\!\Rightarrow|E_{y}\rangle\langle E_{x}|\begin{pmatrix}-1\\
		0\\
		0\\
		1
		\end{pmatrix}=\begin{pmatrix}0\\
		+1\\
		+1\\
		0
		\end{pmatrix}\Longrightarrow & \boxed{\begin{aligned} & "E_{y}"=\\
			& {\textstyle \frac{1}{2}}(\hat{S}_{x}\hat{S}_{y}+\hat{S}_{y}\hat{S}_{x})
			\end{aligned}
		}\\
		|E_{y}\rangle\langle E_{y}|=P_{22}^{(E)}=\frac{2}{6}\!\left[\sum_{n}^{6}(\mathbf{M}_{n})_{22}\hat{O}_{n}\right]= & \frac{1}{2}\!\begin{pmatrix}0\\
		& +1 & +1\\
		& +1 & +1\\
		&  &  & 0
		\end{pmatrix} &  & \qquad\boxed{\begin{aligned} & "E_{x}"=\hat{S}_{x}\hat{S}_{z}+\hat{S}_{z}\hat{S}_{x}\\
			& "E_{y}"=\hat{S}_{y}\hat{S}_{z}+\hat{S}_{z}\hat{S}_{y}
			\end{aligned}
		} & \boxed{\begin{aligned} & "E_{x}"=-\hat{\sigma}_{z}\\
			& "E_{y}"=+\hat{\sigma}_{x}
			\end{aligned}
		}
		\end{aligned}
		\end{equation}
	\end{widetext}	
	where one can observe that $(M_{n}^{(A_1)})_{11}$ (or $(M_{n}^{(A_2)})_{11}$) is simply a number and can therefore be replaced with $\chi_{n}^{(A_{1})}$ (or $\chi_{n}^{(A_{2})}$), which represents the character for the $A_1$ (or $A_2$) irreducible representation in the character table of $C_{3v}$.
	However, since the character table only provides the trace of $(M_{n}^{(E)})_{ij} = (\mathbf{M}_{n})_{ij}$ matrices for two- or higher-dimensional irreducible representations, one can refer to Eq.~\eqref{eq:transformation} for their explicit forms.
	Ultimately, we construct the four symmetrized bases, $"A_1"$, $"A_2"$, $"E_x"$, and $"E_y"$, for a generic polynomial of the form $a\hat{S}_{x}^2+b\hat{S}_{x}\hat{S}_{y}+c\hat{S}_{y}\hat{S}_{x}+d\hat{S}_{y}^{2}$.
	We define $"E_y"$ to follow the right-handed coordinate system established by the $x$, $y$, $z$ bases. Thus, we use the $|E_{y}\rangle\langle E_{x}|$ generalized projector to transform the $"E_x"$ basis into $"E_y"$.
	This approach prevents an unintended projection onto a left-handed coordinate system, which could occur if we relied on $|E_{y}\rangle\langle E_{y}|$.
	In the last row of Eq.~\eqref{eq:SxSy}, we additionally present the symmetrized bases for the orbital operators $\hat{\sigma}_i$ [corresponding to column II in Eq.~\eqref{eq:transformation}] and for the \{$\hat{S}_{x}$, $\hat{S}_{y}$\}$\otimes \hat{S}_{z}$ $"E \otimes A_2"$ product of spin operators, as depicted in Eq.~\eqref{eq:SzSxy}.
	
	Now, we have all the necessary ingredients to construct the orbital-dependent spin Hamiltonian $\hat{H}_{\text{orb}}$ for the $|^3 E\rangle$ excited state. We will express $\hat{H}_{\text{orb}}$ in the following generic form:
	\begin{equation}
	\begin{split}\hat{H}_{\text{orb}}= & -2D_{1}\left["E_{x}\otimes E_{x}"+"E_{y}\otimes E_{y}"\right]\\
	& +2D_{2}\left["E_{x}\otimes E_{x}"+"E_{y}\otimes E_{y}"\right]
	\end{split}
	\label{eq:xx_yy}
	\end{equation}
	where we substitute $"E_x"$ and $"E_y"$ with the bases defined in Eq.~\eqref{eq:SzSxy}. The two arbitrary coefficients are chosen as  $-2D_{1}$ and $+2D_{2}$ to match the Hamiltonian in Eq.\ (3) of Ref.~\onlinecite{Doherty_2013}. We deliberately incorporate the spin-orbit coupling term $-\lambda^{(e)}\hat{S}_{z}\otimes\hat{\sigma}_{y}$ and ground state zero-field splitting terms into Eq.~\eqref{eq:xx_yy} to obtain the final form presented below:
	\begin{equation} 
	\begin{split}\hat{H}^{(e)} & =\hat{H}_{\text{0}}^{(e)}+\hat{H}_{\text{so}}^{(e)}+\hat{H}_{\text{orb}}^{(e)}\\
	& =D^{(e)}\left[\hat{S}_{z}^{2}-\frac{1}{3}S(S+1)\right]\otimes\hat{\sigma}_{0}-\lambda^{(e)}\hat{S}_{z}\otimes\hat{\sigma}_{y}\\
	& \begin{aligned}+D_{1}^{(e)} & \Bigl[ & \!\!\!\!(\hat{S}_{x}\hat{S}_{z}+\hat{S}_{z}\hat{S}_{x}\otimes\hat{\sigma}_{z} & - & \!\!\!\!(\hat{S}_{y}\hat{S}_{z}+\hat{S}_{z}\hat{S}_{y})\otimes\hat{\sigma}_{x}\Bigr]\\
	+D_{2}^{(e)} & \Bigl[ & \!\!\!\!(\hat{S}_{y}^{2}-\hat{S}_{x}^{2})\otimes\hat{\sigma}_{z} & - & \!\!\!\!(\hat{S}_{x}\hat{S}_{y}+\hat{S}_{y}\hat{S}_{x})\otimes\hat{\sigma}_{x}\Bigr]\text{,}
	\end{aligned}
	\end{split}
	\label{eq:Doherty_Spin}
	\end{equation}
	which agrees Eq.\ (3) of Ref.~\onlinecite{Doherty_2013}. One may freely substitute the electron spin operators $\hat{S}_i$ with the nuclear spin operators $\hat{I}_i$ to obtain the excited state Hamiltonian $\hat{H}^{(e)}$, which governs the motion of $^{14}$N nuclear spins.  This Hamiltonian includes nuclear quadrupole interactions ($P^{(e)}$, $P_{1}^{(e)}$, $P_{2}^{(e)}$) and hyperfine interactions ($A^{(e)}_\parallel$, $A^{(e)}_\perp$, $A_{1}^{(e)}$, $A_{2}^{(e)}$). Thus, we present the full excited-state Hamiltonian, including strain and Zeeman terms, but without the effects of $^{13}$C nuclear spins: 
	\begin{equation} 
	\begin{split}\hat{H}^{(e)} & =\overset{\text{Zeeman}}{\overbrace{\mu_{B}g_{S}\hat{\boldsymbol{S}}\boldsymbol{B}+\mu_{B}g_{L}\hat{\sigma}_{y}B_{z}-\mu_{I}g_{I}\hat{\boldsymbol{I}}\boldsymbol{B}}}-\overset{\text{Strain}}{\overbrace{\delta_{x}\hat{\sigma}_{z}+\delta_{y}\hat{\sigma}_{x}}}\\
	& +\underset{\text{Spin-spin zero-field splitting}}{\underbrace{D^{(e)}\left[\hat{S}_{z}^{2}-\frac{1}{3}S(S+1)\right]}}-\underset{\text{spin-orbit}}{\underbrace{\lambda^{(e)}\hat{S}_{z}\hat{\sigma}_{y}}}\\
	& \underset{\text{orbital dependent spin-spin zero-field splitting}}{\underbrace{\begin{aligned}+D_{1}^{(e)} & \Bigl[ & \!\!\!\!(\hat{S}_{x}\hat{S}_{z}+\hat{S}_{z}\hat{S}_{x})\hat{\sigma}_{z} & - & \!\!\!\!(\hat{S}_{y}\hat{S}_{z}+\hat{S}_{z}\hat{S}_{y})\hat{\sigma}_{x}\Bigr]\\
			+D_{2}^{(e)} & \Bigl[ & \!\!\!\!(\hat{S}_{y}^{2}-\hat{S}_{x}^{2})\hat{\sigma}_{z} & - & \!\!\!\!(\hat{S}_{x}\hat{S}_{y}+\hat{S}_{y}\hat{S}_{x})\hat{\sigma}_{x}\Bigr]
			\end{aligned}
	}}\\
	& +\underset{\text{nuclear quadrupole interaction}}{\underbrace{P^{(e)}\left[\hat{I}_{z}^{2}-\frac{1}{3}I(I+1)\right]}\qquad}-\underset{\text{orbital hyperfine}}{\underbrace{A_{\text{orb}}^{(e)}\hat{I}_{z}\otimes\hat{\sigma}_{y}}}\\
	& \underset{\text{orbital dependent nuclear quadrupole interaction}}{\underbrace{\begin{aligned}+P_{1}^{(e)} & \Bigl[ & \!\!\!\!(\hat{I}_{x}\hat{I}_{z}+\hat{I}_{z}\hat{I}_{x})\hat{\sigma}_{z} & - & \!\!\!\!(\hat{I}_{y}\hat{I}_{z}+\hat{I}_{z}\hat{I}_{y})\hat{\sigma}_{x}\Bigr]\\
			+P_{2}^{(e)} & \Bigl[ & \!\!\!\!(\hat{I}_{y}^{2}-\hat{I}_{x}^{2})\hat{\sigma}_{z} & - & \!\!\!\!(\hat{I}_{x}\hat{I}_{y}+\hat{I}_{y}\hat{I}_{x})\hat{\sigma}_{x}\Bigr]
			\end{aligned}
	}}\\
	& \underset{\text{hyperfine interaction}}{\underbrace{+\Bigl[A_{\parallel}^{(e)}\hat{S}_{z}\hat{I}_{z}-A_{\perp}^{(e)}(\hat{S}_{x}\hat{I}_{x}+\hat{S}_{y}\hat{I}_{y})\Bigr]}}\\
	& \underset{\text{orbital dependent hyperfine interaction}}{\underbrace{\begin{aligned}+A_{1}^{(e)} & \Bigl[ & \!\!\!\!(\hat{S}_{x}\hat{I}_{z}+\hat{S}_{z}\hat{I}_{x})\hat{\sigma}_{z} & - & \!\!\!\!(\hat{S}_{y}\hat{I}_{z}+\hat{S}_{z}\hat{I}_{y})\hat{\sigma}_{x}\Bigr]\\
			+A_{2}^{(e)} & \Bigl[ & \!\!\!\!(\hat{S}_{y}\hat{I}_{y}-\hat{S}_{x}\hat{I}_{x})\hat{\sigma}_{z} & - & \!\!\!\!(\hat{S}_{x}\hat{I}_{y}+\hat{S}_{y}\hat{I}_{x})\hat{\sigma}_{x}\Bigr] \text{,}
			\end{aligned}
	}}
	\end{split}
	\label{eq:FullHamilton}
	\end{equation}
	where $\mu_{B}$ and $g_S$ are the Bohr magneton and the electron spin $g$-factor, respectively, and $g_L$ is the orbital $g$-factor for the $|e_\pm\rangle$ hole orbital. Similarly, $\mu_{I}$ and $g_I$ represent the nuclear magneton and $g$-factor for the $^{14}$N nuclear spin, respectively. The terms $\delta_x$ and $\delta_y$ describe the splitting of the $|e_x\rangle$, $|e_y\rangle$ orbitals due to external strain or an electric field~\cite{Doherty_2013}. In our specific case, the NV center experienced a strain of approximately $\delta_\perp = \sqrt{\delta_x^2+\delta_y^2} \approx 3 $~GHz, as reported in Ref.~\onlinecite{Monge_2023}. Additionally, we account for the orbital hyperfine interaction~\cite{Blugel_1987, van_Lenthe_1998, Tissot_2021, Franzke_2024}, represented as $A_{\text{orb}}^{(e)}\hat{I}_{z}\otimes\hat{\sigma}_{y}$, where we estimate $A_{\text{orb}}^{(e)}$ to be in the same order of magnitude ($\sim$MHz) as the hyperfine interactions $A^{(e)}_\parallel$ and $A^{(e)}_\perp$. However, we note that the orbital hyperfine interaction can be omitted from further discussion, as the strain splitting $\delta_\perp \approx 3 $~GHz significantly exceeds the $A_{\text{orb}}^{(e)}\sim$~MHz range.  This effectively quenches the orbital moment associated with the $|E_{x,y}\rangle$ $m_S=0$ substates in the $|^3 E\rangle$ optical excited state. Furthermore, it does not induce nuclear spin flips, which are the primary focus of this manuscript.
	
	\section{Spin Hamiltonian in the ladder operator form}
	\label{app:B}
	In accordance with the Condon-Shortley convention, we introduce the circularly polarized "\textit{complex basis}" as follows: $|e_{\pm}\rangle=\mp\bigl(|e_{x}\rangle\pm i|e_{y}\rangle\bigr)/\sqrt{2}$.
	Thus, the basis transformation can be represented by the matrix $\mathbf{V}=\bigl(\begin{smallmatrix}-1 & -i\\
	+1 & -i
	\end{smallmatrix}\bigr)/\sqrt{2}$ since $\bigl(\begin{smallmatrix}|e_{+}\rangle\\
	|e_{-}\rangle
	\end{smallmatrix}\bigr)=\mathbf{V}\bigl(\begin{smallmatrix}|e_{x}\rangle\\
	|e_{y}\rangle
	\end{smallmatrix}\bigr)$. Therefore, the transformation matrices for the $E$ irreducible representation in the "\textit{angular}" basis \{$E_+$, $E_-$\}, represented by $\bigl(\begin{smallmatrix}|e_{+}\rangle\\
	|e_{-}\rangle
	\end{smallmatrix}\bigr)$, are given by $\mathbf{N}_{n}=\mathbf{V}\mathbf{M}_{n}\mathbf{V}^{-1}$ as shown in Eq.~\eqref{eq:complexRepresent} below (for further details, see Table T~\textbf{51}.7\textit{A} in the book by Altmann and Herzig~\cite{Altmann_2011}):
	\begin{equation}
	\begin{aligned} & \underset{{\textstyle \hat{E}}}{\underbrace{\mathbf{N}_{1}=\bigl(\begin{smallmatrix}1 & 0\\
			0 & 1
			\end{smallmatrix}\bigr)}}\text{,} &  & \underset{{\textstyle \hat{C}_{3}^{+}}}{\underbrace{\mathbf{N}_{2}=\bigl(\begin{smallmatrix}\varepsilon^{*} & 0\\
			0 & \varepsilon
			\end{smallmatrix}\bigr)}}\text{,} &  & \underset{{\textstyle \hat{C}_{3}^{-}}}{\underbrace{\mathbf{N}_{3}=\bigl(\begin{smallmatrix}\varepsilon & 0\\
			0 & \varepsilon^{*}
			\end{smallmatrix}\bigr)}}\text{,}\\
	& \underset{{\textstyle \hat{\sigma}_{v}^{(1)}}}{\underbrace{\mathbf{N}_{4}=\bigl(\begin{smallmatrix}0 & -1\\
			-1 & 0
			\end{smallmatrix}\bigr)}}\text{,} &  & \underset{{\textstyle \hat{\sigma}_{v}^{(2)}}}{\underbrace{\mathbf{N}_{5}=\bigl(\begin{smallmatrix}0 & -\varepsilon\\
			-\varepsilon^{*} & 0
			\end{smallmatrix}\bigr)}}\text{,} &  & \underset{{\textstyle \hat{\sigma}_{v}^{(3)}}}{\underbrace{\mathbf{N}_{6}=\bigl(\begin{smallmatrix}0 & -\varepsilon^{*}\\
			-\varepsilon & 0
			\end{smallmatrix}\bigr)}}\text{,}
	\end{aligned}
	\label{eq:complexRepresent}
	\end{equation}
	where $\varepsilon = e^{+2\pi i/3 }$ and $\varepsilon^* = e^{-2\pi i/3 }$, thus $\hat{E}\bigl(\begin{smallmatrix}|e_{+}\rangle\\
	|e_{-}\rangle
	\end{smallmatrix}\bigr)=\mathbf{N}_{1}\bigl(\begin{smallmatrix}|e_{+}\rangle\\
	|e_{-}\rangle
	\end{smallmatrix}\bigr)$, ... $\hat{\sigma}_{v}^{(3)}\bigl(\begin{smallmatrix}|e_{+}\rangle\\
	|e_{-}\rangle
	\end{smallmatrix}\bigr)=\mathbf{N}_{6}\bigl(\begin{smallmatrix}|e_{+}\rangle\\
	|e_{-}\rangle
	\end{smallmatrix}\bigr)$. The advantage here is that the $\hat{C}_3^{\pm}$ rotations remain diagonal in the "\textit{complex basis}," in contrast to their representation in column IV of Eq.~\eqref{eq:transformation} for $\mathbf{M}_{n}$s in the "\textit{real valued}" basis. Thus, one may convert the orbital operators \{$-\hat{\sigma}_z $, $\hat{\sigma}_x $\}, $\hat{\sigma}_y $ into their ladder operator form, $\hat{\sigma}_\mp=|e_{\mp}\rangle\langle e_{\pm}|$:
	\begin{equation}
	\begin{aligned}\begin{aligned}-\hat{\sigma}_{z}=|e_{y}\rangle\langle e_{y}|-|e_{x}\rangle\langle e_{x}|=\underset{\qquad\qquad{\textstyle =\hat{\sigma}_{-}+\hat{\sigma}_{+}\text{,}}}{|e_{-}\rangle\langle e_{+}|+|e_{+}\rangle\langle e_{-}|}\end{aligned}
	\\
	\begin{aligned}\hat{\sigma}_{x}=|e_{x}\rangle\langle e_{y}|+|e_{y}\rangle\langle e_{x}|=\underset{{\textstyle \qquad\qquad=i(\hat{\sigma}_{-}-\hat{\sigma}_{+})\text{,}}}{i(|e_{-}\rangle\langle e_{+}|-|e_{+}\rangle\langle e_{-}|)}\end{aligned}
	\\
	\begin{aligned}\hat{\sigma}_{y}=i(|e_{y}\rangle\langle e_{x}|-|e_{x}\rangle\langle e_{y}|)=|e_{+}\rangle\langle e_{+}|-|e_{-}\rangle\langle e_{-}|=\hat{\sigma}_{z}\text{.}\end{aligned}
	\end{aligned}
	\label{eq:operators_complex}
	\end{equation}
	Thus, one may use the following transformation rules to convert $\hat{H}_{\text{orb}}$ from Eq.~\eqref{eq:Doherty_Spin} into the "\textit{complex basis}"
	\begin{equation}
	\begin{aligned} & \mathbf{V}=\frac{1}{\sqrt{2}}\begin{pmatrix}-1 & -i\\
	+1 & -i
	\end{pmatrix}\text{,} & \quad & \Bigl(\begin{smallmatrix}|e_{x}\rangle\\
	|e_{y}\rangle
	\end{smallmatrix}\Bigr)=\mathbf{V}\Bigl(\begin{smallmatrix}|e_{x}\rangle\\
	|e_{y}\rangle
	\end{smallmatrix}\Bigr)\text{,}\\
	& \bigl(\begin{smallmatrix}-\hat{\sigma}_{-}\\
	+\hat{\sigma}_{+}
	\end{smallmatrix}\bigr)=\frac{1}{\sqrt{2}}\mathbf{V}\bigl(\begin{smallmatrix}-\hat{\sigma}_{z}\\
	+\hat{\sigma}_{x}
	\end{smallmatrix}\bigr)\text{,} & \quad & \Bigl(\begin{smallmatrix}-\hat{S}_{+}\\
	+\hat{S}_{-}
	\end{smallmatrix}\Bigr)=\sqrt{2}\mathbf{V}\Bigl(\begin{smallmatrix}\hat{S}_{x}\\
	\hat{S}_{y}
	\end{smallmatrix}\Bigr)\text{,}
	\end{aligned}
	\label{eq:inverse rules}
	\end{equation}
	and identify the $\bigl(\begin{smallmatrix}"E_{+}"\\
	"E_{-}"
	\end{smallmatrix}\bigr)$ combinations that transform according to $\mathbf{N}_{n}$ matrices in Eq.~\eqref{eq:complexRepresent}. Additionally, we introduce the ladder operators for spin as usual: $\hat{S}_{\pm}=\hat{S}_{x}\pm i\hat{S}_{y}$.
	\begin{equation}
	\begin{split} \hat{H}^{(e)}  = &  \hat{H}_{\text{0}}^{(e)}+\hat{H}_{\text{so}}^{(e)}+\hat{H}_{\text{orb}}^{(e)} \\ = & D\left[\hat{S}_{z}^{2}-\frac{1}{3}S(S+1)\right]-\lambda\hat{L}_{z}\hat{S}_{z}\\
	- & D_{1}\left[(\hat{S}_{-}\hat{S}_{z}+\hat{S}_{z}\hat{S}_{-})\hat{\sigma}_{-}+(\hat{S}_{+}\hat{S}_{z}+\hat{S}_{z}\hat{S}_{+})\hat{\sigma}_{+}\right]\\
	+ & D_{2}\left[\hat{S}_{+}^{2}\hat{\sigma}_{-}+\hat{S}_{-}^{2}\hat{\sigma}_{+}\right]
	\end{split}
	\label{eq:Hamiltonian_ladder}
	\end{equation}
	The orbital operators involving the $|a_{1}\rangle$ level are shown below:
	\begin{equation}
	\hat{\sigma}_{\mp}|e_{\pm}\rangle=\frac{1}{2}\hat{L}_{\mp}^{2}|e_{\mp}\rangle=\frac{1}{\sqrt{2}}\hat{L}_{\mp}|a_{1}\rangle=|e_{\mp}\rangle
	\end{equation}
	
	We evaluate the matrix elements of the excited state Hamiltonian [Eq.~\eqref{eq:Hamiltonian_ladder}] using the $|E_{m_{L}}^{m_{S}}\rangle=|a_{1}e_{\pm}\rangle\otimes|m_{S}\rangle$ wavefunctions:
	\begin{widetext}
		\begin{equation}
		\langle E_{m_{L}}^{m_{S}}|\hat{H}^{(e)}|E_{m_{L}}^{m_{S}}\rangle=\left(\begin{array}{cccccc}
		\frac{2}{3}D^{(e)}+\lambda_{z}^{(e)} & 2D_{2}^{(e)} & 0 & 0 & 0 & 0\\
		2D_{2}^{(e)} & \frac{2}{3}D^{(e)}+\lambda_{z}^{(e)} & 0 & 0 & 0 & 0\\
		0 & 0 & -\frac{1}{3}D^{(e)} & 0 & \sqrt{2}D_{1}^{(e)} & 0\\
		0 & 0 & 0 & -\frac{1}{3}D^{(e)} & 0 & -\sqrt{2}D_{1}^{(e)}\\
		0 & 0 & \sqrt{2}D_{1}^{(e)} & 0 & \frac{2}{3}D^{(e)}-\lambda_{z}^{(e)} & 0\\
		0 & 0 & 0 & -\sqrt{2}D_{1}^{(e)} & 0 & \frac{2}{3}D^{(e)}-\lambda_{z}^{(e)}
		\end{array}\right)\begin{array}{cc}
		\leftarrow & |E_{-}^{\uparrow\uparrow}\rangle\\
		\leftarrow & |E_{+}^{\downarrow\downarrow}\rangle\\
		\leftarrow & |E_{+}^{0}\rangle\\
		\leftarrow & |E_{-}^{0}\rangle\\
		\leftarrow & |E_{-}^{\downarrow\downarrow}\rangle\\
		\leftarrow & |E_{+}^{\uparrow\uparrow}\rangle
		\end{array}
		\text{,}
		\label{eq:matrixform}
		\end{equation}
	\end{widetext}
	where 
	$\bigl(\begin{smallmatrix}|E_{x}\rangle\\
	|E_{y}\rangle
	\end{smallmatrix}\bigr)$ and $\bigl(\begin{smallmatrix}|E_{1}\rangle\\
	|E_{2}\rangle
	\end{smallmatrix}\bigr)$ transform as $\mathbf{M}_{1}$ matrices in Eq.~\eqref{eq:transformation}. $|m_{S}\rangle=\{|{\uparrow\uparrow}\rangle,(|{\uparrow\downarrow}\rangle+|{\downarrow\uparrow}\rangle)/\sqrt{2},|{\downarrow\downarrow}\rangle\}$ and $|E_{\pm}\rangle=(|a_{1}e_{\pm}\rangle-|e_{\pm}a_{1}\rangle)/\sqrt{2}$ describe the orbital and spin parts of the two-hole wavefunction of the $|^3 E\rangle$ manifold. 
	We utilize the following Clebsch-Gordan coefficients of the $C_{3v}$ double point group from Table $\mathbf{51}$.11\textit{A} of Ref.~\onlinecite{Altmann_2011}:  $(E_{+}\otimes E_{-}+E_{-}\otimes E_{+})/\sqrt{2}=A_{1}$; $(E_{+}\otimes E_{-}-E_{-}\otimes E_{+})/\sqrt{2}=A_{2}$; $E_{+}\otimes E_{+}=-E_{-}$; $E_{-}\otimes E_{-}=+E_{+}$; ${}\uparrow\otimes{}\uparrow=+E_{+}$; 
	${}\downarrow{}\otimes{}\downarrow{}=-E_{-}$. These allow us to define the following spin-orbit states:
	\begin{equation}
	\begin{alignedat}{6}|A_{1}\rangle & = & {\textstyle \frac{1}{\sqrt{2}}}(|E_{-}^{\uparrow\uparrow}\rangle & \,-\, & |E_{+}^{\downarrow\downarrow}\rangle)\,\text{,} & \;\; & |A_{2}\rangle & = & {\textstyle \frac{1}{\sqrt{2}}}(|E_{-}^{\uparrow\uparrow}\rangle & \,+\, & |E_{+}^{\downarrow\downarrow}\rangle)\,\text{,}\\
	|E_{x}\rangle & = & {\textstyle \frac{-1}{\sqrt{2}}}(|E_{+}^{0}\rangle & \,+\, & |E_{-}^{0}\rangle)\,\text{,} &  & |E_{y}\rangle & = & {\textstyle \frac{i}{\sqrt{2}}}(|E_{+}^{0}\rangle & \,-\, & |E_{-}^{0}\rangle)\,\text{,}\\
	|E_{1}\rangle & = & {\textstyle \frac{1}{\sqrt{2}}}(|E_{-}^{\downarrow\downarrow}\rangle & \,-\, & |E_{+}^{\uparrow\uparrow}\rangle)\,\text{,} &  & |E_{2}\rangle & = & {\textstyle \frac{-i}{\sqrt{2}}}(|E_{-}^{\downarrow\downarrow}\rangle & \,+\, & |E_{+}^{\uparrow\uparrow}\rangle)\,\text{,}
	\end{alignedat}
	\label{eq:SO_multiplets}
	\end{equation}
	in which $\bigl(\begin{smallmatrix}|E_{x}\rangle\\
	|E_{y}\rangle
	\end{smallmatrix}\bigr)$ and $\bigl(\begin{smallmatrix}|E_{1}\rangle\\
	|E_{2}\rangle
	\end{smallmatrix}\bigr)$ transform as $\mathbf{M}_{1}$, $\mathbf{M}_{2}$, ...  $\mathbf{M}_{6}$ matrices in Eq.~\eqref{eq:transformation}. Thus, in simple terms, they transform similarly to the $(\begin{smallmatrix}x\\
	y
	\end{smallmatrix}\bigr)$ coordinates shown in Fig.~\ref{fig:transform}. The excited state Hamiltonian, expressed in the basis defined in Eq.~\eqref{eq:SO_multiplets}, is given as follows:
	\begin{equation}
	\begin{alignedat}{1} & \hat{H}^{(e)}=  -\frac{1}{3}D^{(e)}\Bigl(|E_{x}\rangle\langle E_{x}|+|E_{y}\rangle\langle E_{y}|\Bigr)\\
	& +\frac{2}{3}D^{(e)}\Bigl(|A_{1}\rangle\langle A_{1}|+|A_{2}\rangle\langle A_{2}|+|E_{1}\rangle\langle E_{1}|+|E_{2}\rangle\langle E_{2}|\Bigr)\\
	& +\lambda_{z}^{(e)}\Bigl(|A_{1}\rangle\langle A_{1}|+|A_{2}\rangle\langle A_{2}|-|E_{1}\rangle\langle E_{1}|-|E_{2}\rangle\langle E_{2}|\Bigr)\\
	& +\sqrt{2}D_{1}^{(e)}\Bigl(|E_{1}\rangle\langle E_{x}|+|E_{2}\rangle\langle E_{y}|+|E_{x}\rangle\langle E_{1}|+|E_{2}\rangle\langle E_{y}|\Bigr)  \\
	& +D_{2}^{(e)}\Bigl(|A_{1}\rangle\langle A_{1}|-|A_{2}\rangle\langle A_{2}|\Bigr) \text{    ,} 
	\end{alignedat}
	\end{equation}
	which is in agreement with Tables~2 and 3 in Ref.~\onlinecite{Doherty_2011}.

	\section{Literature review on the $C_{3v}$ double group}
	\label{app:C}
	In analyzing the spin Hamiltonian of the $^3E$ excited state, we recognized that different conventions employed in the literature~\cite{Doherty_2011, Maze2011} may cause confusion for readers, particularly regarding the definition of orbitals or operators. Furthermore, we identified typographical errors in the reported definitions within the given conventions and in the final spin Hamiltonian operator. For clarity, we list these instances below.

	\subsection{Different conventions}
	\label{app:Ca}
	
	The definition of $|^3E \rangle$ manifold depends on the applied convention. Our paper uses the same convention as Ref.~\onlinecite{Doherty_2011} but distinct from that of Ref.~\onlinecite{Maze2011} (see Table~1 there). In the latter, the definition is
	\begin{equation}
	\begin{alignedat}{6}|A_{1}\rangle & = & {\textstyle \frac{1}{\sqrt{2}}}(|E_{-}^{\uparrow\uparrow}\rangle & \,-\, & |E_{+}^{\downarrow\downarrow}\rangle)\,\text{,} & \quad & |A_{2}\rangle & = & {\textstyle \frac{1}{\sqrt{2}}}(|E_{-}^{\uparrow\uparrow}\rangle & \,+\, & |E_{+}^{\downarrow\downarrow}\rangle)\,\text{,}\\
	\underset{\!"-i\xi_{x}"\!}{\underbrace{|E_{y}\rangle}} & = & {\textstyle \frac{i}{\sqrt{2}}}(|E_{-}^{0}\rangle & \,+\, & |E_{+}^{0}\rangle)\,\text{,} & \quad & \underset{\!"-\xi_{y}"\!}{\underbrace{|E_{x}\rangle}} & = & {\textstyle \frac{1}{\sqrt{2}}}(|E_{-}^{0}\rangle & \,-\, & |E_{+}^{0}\rangle)\,\text{,}\\
	\underset{\!"+\xi_{x}"\!}{\underbrace{|E_{1}\rangle}} & = & {\textstyle \frac{1}{\sqrt{2}}}(|E_{-}^{\downarrow\downarrow}\rangle & \,-\, & |E_{+}^{\uparrow\uparrow}\rangle)\,\text{,} & \quad & \underset{\!\!\!"+i\xi_{y}"\!}{\underbrace{|E_{2}\rangle}} & = & {\textstyle \frac{1}{\sqrt{2}}}(|E_{-}^{\downarrow\downarrow}\rangle & \,+\, & |E_{+}^{\uparrow\uparrow}\rangle)\,\text{,}
	\end{alignedat}
	\end{equation}
	where $-i\xi_{x}$, $-\xi_{y}$, $+\xi_{x}$, $i\xi_{y}$ indicate that the $|E_{x,y,1,2}\rangle$ states in this form do not follow the $\mathbf{M}_{n}$ matrices from Eq.~\eqref{eq:transformation} only the $\{\frac{|E_{y}\rangle}{-i},\frac{|E_{x}\rangle}{-1}\}$ and $\{\frac{|E_{1}\rangle}{+1},\frac{|E_{2}\rangle}{+i}\}$ combinations are transforming as $\{"x","y"\}$ vectors (see Fig.~\ref{fig:transform}) thus following $\mathbf{M}_{n}$ matrices. Nevertheless, the matrix elements of the $\hat{H}^{(e)}$ Hamiltonian can still be calculated in this basis:
	\begin{equation}
	\begin{alignedat}{1} & \hat{H}_{\mathrm{ours}}^{(e)}=-\frac{1}{3}D^{(e)}\Bigl(|E_{x}\rangle\langle E_{x}|+|E_{y}\rangle\langle E_{y}|\Bigr)\\
	& +\frac{2}{3}D^{(e)}\Bigl(|A_{1}\rangle\langle A_{1}|+|A_{2}\rangle\langle A_{2}|+|E_{1}\rangle\langle E_{1}|+|E_{2}\rangle\langle E_{2}|\Bigr)\\
	& +\lambda_{z}^{(e)}\Bigl(|A_{1}\rangle\langle A_{1}|+|A_{2}\rangle\langle A_{2}|-|E_{1}\rangle\langle E_{1}|-|E_{2}\rangle\langle E_{2}|\Bigr)\\
	& \!\!+\!\sqrt{2}D_{1}^{(e)}\!\Bigl(-i|E_{1}\rangle\langle E_{y}|\!+\!i|E_{y}\rangle\langle E_{1}|\!+\!|E_{2}\rangle\langle E_{x}|\!+\!|E_{x}\rangle\langle E_{2}|\Bigr)\\
	& +2D_{2}^{(e)}\Bigl(|A_{2}\rangle\langle A_{2}|-|A_{1}\rangle\langle A_{1}|\Bigr)
	\end{alignedat}
	\end{equation}
	which differs from Eq.~(8) in Ref.~\onlinecite{Maze2011} that we highlight below:
	\begin{equation}
	\begin{alignedat}{1} & \hat{H}_{\mathrm{Maze}}^{(e)}=-\frac{1}{3}D^{(e)}\Bigl(|E_{x}\rangle\langle E_{x}|+|E_{y}\rangle\langle E_{y}|\Bigr)\\
	& +\frac{2}{3}D^{(e)}\Bigl(|A_{1}\rangle\langle A_{1}|+|A_{2}\rangle\langle A_{2}|+|E_{1}\rangle\langle E_{1}|+|E_{2}\rangle\langle E_{2}|\Bigr)\\
	& +\lambda_{z}^{(e)}\Bigl(|A_{1}\rangle\langle A_{1}|+|A_{2}\rangle\langle A_{2}|-|E_{1}\rangle\langle E_{1}|-|E_{2}\rangle\langle E_{2}|\Bigr)\\
	& +\boxed{D_{1}^{(e)}\Bigl(|E_{1}\rangle\langle E_{y}|\!+\!|E_{y}\rangle\langle E_{1}|\!-\!i|E_{2}\rangle\langle E_{x}|\!+\!i|E_{x}\rangle\langle E_{2}|\Bigr)}\\
	& +2D_{2}^{(e)}\Bigl(|A_{2}\rangle\langle A_{2}|-|A_{1}\rangle\langle A_{1}|\Bigr)
	\end{alignedat}
	\end{equation}
	Notice that the complex unit $i$ is incorrectly positioned. Additionally, the extra factor of $\sqrt{2}$ may originate from a different notation for spin-spin interaction.
	
	\subsection{Definition of transformation matrices}
	\label{app:Cb}
	
	The transformation matrices reported in Table~14 of Ref.~\onlinecite{Doherty_2013} should be revisited. The matrices are presented as follows:
	\begin{equation}
	\begin{aligned} & \underset{{\textstyle \hat{E}}}{\underbrace{\mathbf{N}_{1}=\bigl(\begin{smallmatrix}1 & 0\\
			0 & 1
			\end{smallmatrix}\bigr)}}\text{,} &  & \underset{{\textstyle \hat{C}_{3}^{+}}}{\underbrace{\mathbf{N}_{2}=\bigl(\begin{smallmatrix}\boxed{\varepsilon} & 0\\
			0 & \boxed{\varepsilon^{*}}
			\end{smallmatrix}\bigr)}}\text{,} &  & \underset{{\textstyle \hat{C}_{3}^{-}}}{\underbrace{\mathbf{N}_{3}=\bigl(\begin{smallmatrix}\boxed{\varepsilon^{*}} & 0\\
			0 & \boxed{\varepsilon}
			\end{smallmatrix}\bigr)}}\text{,}\\
	& \underset{{\textstyle \hat{\sigma}_{v}^{(1)}}}{\underbrace{\mathbf{N}_{4}=\bigl(\begin{smallmatrix}0 & -1\\
			-1 & 0
			\end{smallmatrix}\bigr)}}\text{,} &  & \underset{{\textstyle \hat{\sigma}_{v}^{(2)}}}{\underbrace{\mathbf{N}_{5}=\bigl(\begin{smallmatrix}0 & -\varepsilon\\
			-\varepsilon^{*} & 0
			\end{smallmatrix}\bigr)}}\text{,} &  & \underset{{\textstyle \hat{\sigma}_{v}^{(3)}}}{\underbrace{\mathbf{N}_{6}=\bigl(\begin{smallmatrix}0 & -\varepsilon^{*}\\
			-\varepsilon & 0
			\end{smallmatrix}\bigr)}\text{,}}
	\end{aligned}
	\label{eq:Doherty:bad}
	\end{equation}
	which results in erroneous phases when compared to our results in Eq.~\eqref{eq:complexRepresent} or those presented in the book by Altmann and Herzig~\cite{Altmann_2011}. 
	
	\subsection{Sign of zero-field splitting}
	\label{app:Cc}
	
	We note that Ref.~\onlinecite{Doherty_2011} originally depicted the highest energy state as $|A_1\rangle$, followed by $|A_2\rangle$. However, a later correction in Ref.~\onlinecite{Doherty_2013} established that $|A_2\rangle$ is actually the highest energy state, a conclusion that has since been adopted in subsequent works by various authors. Our \textit{ab initio} calculations further validate this correction by determining the proper energy ordering of $|A_1\rangle$ and $|A_2\rangle$ and confirming that the sign of $D_2$ is positive, thereby reaffirming that $|A_2\rangle$ is indeed the uppermost state.
	
	\section{Discussion of the JT reduction factors}
	\label{app:D}
	
	A known relationship between JT reduction factors $p$ and $q$ yields $q=(1+p)/2$ for $E \otimes e$ JT system.
	This equality strictly holds only for a linear JT effect coupled to a single $E$ vibrational mode. However, it remains accurate within a few percent even when accounting for quadratic JT terms or multiple phonon modes. Incorporating quadratic terms---arising from JT barrier energies~\cite{Halperin_1969, Slonczewski_1969})---or considering multiple phonon modes (multimode problem~\cite{Polinger_1979, Polinger_1979b, Hoffman_1983}) results in $q<(1+p)/2$.
	However, the deviation is within a few percent: (i) the ratio between $q$ and $q_0$ is negligible for multimode $E\otimes (e\oplus e \oplus...\oplus e)$ problems at $k_\text{eff}=1$ in Figure~5 of Ref.~\onlinecite{Evangelou_1980}; (ii) Table~I in the Supplemental Materials of Ref.~\onlinecite{Thiering2017} indicates an insignificant correction ($f_1<0.001$) for $q$ in the single-mode quadratic $E\otimes e $ problem of NV center in diamond; additionally, see Eq.~(15) therein and compare to $q=(1+p)/2$ of the main text.  
	Finally, we note that the JT tunneling rate---characterized by a frequency $\sim8.5$~THz~\cite{Abtew2011} and known as \textit{tunneling splitting}~\cite{bersuker1963sov, Bersuker_1975, Bersuker_1975b, Ham_1968, Garcia-Fernandez2012, O_Brien_1993, Garcia_2010, Bersuker_2023}---exceeds all spin interactions discussed here by several orders of magnitude.
	Therefore, $p$ and $q$ can be interpreted as renormalization factors for physical observables~\cite{Norambuena_2020} associated with the $|^3 E\rangle$ excited state, which ultimately behaves as a quasiparticle dressed by electron-phonon interaction. The entanglement induced by the dynamic JT, encoded in $p$ and $q$, can only be disrupted under two conditions: (i) when the system is probed using ultrafast femtosecond spectroscopy~\cite{Ulbricht_2016, Ulbricht_2018, Carbery_2024, Huxter_2013}, or (ii) when the JT tunneling rate across the saddle points [Fig.~\ref{fig:APES}(a)] is sufficiently slow for static JT distortions~\cite{Reynolds_1975, Isoya_1990, bersuker2012vibronic, bersuker2013jahn} to be observed---neither of which applies to the NV center in diamond.

	\section{\textit{Ab initio} treatment of spin Hamiltonian tensors}
	\label{app:E}
	
	Through \textit{ab initio} calculations, one can determine the spin Hamiltonian tensors by assuming the orbital state to be $|e_{\varphi}\rangle=\cos(\frac{\varphi}{2})|e_{x}\rangle-\sin(\frac{\varphi}{2})|e_{y}\rangle$, where $\varphi$ is a rotation angle that can be affected by numerical instabilities. If the JT APES exhibits energy barriers then $\varphi$ stabilizes at one of the three minima: $\varphi=n\times120^{\circ}$ (where $n=0, 1, 2$ is an integer).
	The angles $\varphi=60^{\circ} + n\times120^{\circ}$ correspond to saddle points of the APES, separating the minima by energy barriers.
	
	Now, we determine the non-trivial components of the spin tensors $\boldsymbol{A}$, $\boldsymbol{D}$, $\boldsymbol{P}$.
	Here, we will follow the derivation for $D$, though a similar approach applies to $\boldsymbol{A}$ and $\boldsymbol{P}$ as well.
	If the orbital $|e_{x}\rangle$ ($\varphi=0^{\circ}$) is chosen, then the $\boldsymbol{D}$ tensor takes the form $\boldsymbol{D}^{(xx)}$. Conversely, if the orbital $|e_{y}\rangle$ ($\varphi=180^{\circ}$) is selected, the observed tensor is $\boldsymbol{D}^{(yy)}$.
	As a result, \textit{ab initio} calculations generally yield different $\boldsymbol{D}^{(\varphi)}$ tensors, which may appear to break the $C_{3v}$ symmetry. However, we will show that the off-diagonal terms $\boldsymbol{D}^{(xy)}$ and $\boldsymbol{D}^{(yx)}$ are inaccessible within Kohn-Sham DFT.
	Nonetheless, these terms can still be determined through group theory analysis or by evaluating $\boldsymbol{D}^{(\varphi)}$ at intermediate angles. To systematically analyze the zero-field splitting tensor, we decompose it into components using the Pauli matrices ($\hat{\sigma}_{x}=(\begin{smallmatrix}&1\\1&\end{smallmatrix})$, $\hat{\sigma}_{y}=(\begin{smallmatrix}&\!\!-i\\i&\end{smallmatrix})$, $\hat{\sigma}_{z}=(\begin{smallmatrix}1&\\&\!\!-1\end{smallmatrix})$) for the orbital degrees of freedom
	\begin{widetext}
		\begin{equation}
		\label{eq:SM:1}
		\begin{split}\hat{W}= & \hat{\boldsymbol{S}}^{T}\boldsymbol{D}^{(xx)}\hat{\boldsymbol{S}}|e_{x}\rangle\langle e_{x}|+\hat{\boldsymbol{S}}^{T}\boldsymbol{D}^{(xy)}\hat{\boldsymbol{S}}|e_{x}\rangle\langle e_{y}|+\hat{\boldsymbol{S}}^{T}\boldsymbol{D}^{(yx)}\hat{\boldsymbol{S}}|e_{y}\rangle\langle e_{x}|+\hat{\boldsymbol{S}}^{T}\boldsymbol{D}^{(yy)}\hat{\boldsymbol{S}}|e_{y}\rangle\langle e_{y}|=\\
		& \hat{\boldsymbol{S}}^{T}\boldsymbol{D}^{(xx)}\hat{\boldsymbol{S}}\frac{(1+\hat{\sigma}_{z})}{2}+\hat{\boldsymbol{S}}^{T}\boldsymbol{D}^{(xy)}\hat{\boldsymbol{S}}\frac{(\hat{\sigma}_{x}+i\sigma_{y})}{2}+\hat{\boldsymbol{S}}^{T}\boldsymbol{D}^{(yx)}\hat{\boldsymbol{S}}\frac{(\hat{\sigma}_{x}-i\sigma_{y})}{2}+\hat{\boldsymbol{S}}^{T}\boldsymbol{D}^{(yy)}\hat{\boldsymbol{S}}\frac{(1-\hat{\sigma}_{z})}{2}=\\
		& \underset{1\:\text{transforms as } A_{1}}{\underbrace{\hat{\boldsymbol{S}}^{T}\frac{1}{2}\Bigl[\boldsymbol{D}^{(xx)}+\boldsymbol{D}^{(yy)}\Bigr]\hat{\boldsymbol{S}}}}
		+\\
		& \underset{\text{\{\ensuremath{\hat{\sigma}_{z}};\ensuremath{\hat{\sigma}_{x}}\}\:transforms as } E}{\underbrace{\hat{\boldsymbol{S}}^{T}\frac{1}{2}\Bigl[\boldsymbol{D}^{(xx)}-\boldsymbol{D}^{(yy)}\Bigr]\hat{\boldsymbol{S}}\hat{\sigma}_{z}+
				\hat{\boldsymbol{S}}^{T}\frac{1}{2}\Bigl[\boldsymbol{D}^{(xy)}+\boldsymbol{D}^{(yx)}\Bigr]\hat{\boldsymbol{S}}\hat{\sigma}_{x}}}+\underset{i\hat{\sigma}_{y}=
			L_{z}\:\text{transforms as } A_{2}}{\underbrace{\hat{\boldsymbol{S}}^{T}\frac{1}{2}\Bigl[\boldsymbol{D}^{(xy)}-\boldsymbol{D}^{(yx)}\Bigr]\hat{\boldsymbol{S}}i\hat{\sigma}_{y}}} \text{,}
		\end{split}
		\end{equation}
	\end{widetext}
	where the chosen decomposition highlights the symmetry of the orbital operators, with $\hat{\boldsymbol{S}}^{T}=(\hat{S}_{x}\;\hat{S}_{y}\;\hat{S}_{z})$.
	The first symmetrized term transforms as the $A_{1}$ irreducible representation, thereby containing the $D$ parameter.
	Additionally, two terms transform as the $E$ irreducible representation of $C_{3v}$, denoted as $D_{1}$ and $D_{2}$.
	However, it is important to note that no terms transform as $A_{2}$ because all tensors are symmetric, allowing us to neglect the last term.
	
	Simultaneously, we construct the symmetry-adapted $\hat{H}_\text{orb}$ interaction terms, incorporating second-order $\hat{S}_i$ operators as follows:
	\begin{equation}
	\label{eq:SM:2}
	\begin{split}\hat{H}_\text{orb} = & D\left[\hat{S}_{z}^{2}-\frac{1}{3}S(S+1)\right]+\\
	& D_{1}\left[(\hat{S}_{x}\hat{S}_{z}+\hat{S}_{z}\hat{S}_{x})\hat{\sigma}_{z}-(\hat{S}_{y}\hat{S}_{z}+\hat{S}_{z}\hat{S}_{y})\hat{\sigma}_{x}\right]+\\
	& D_{2}\left[(\hat{S}_{y}\hat{S}_{y}-\hat{S}_{x}\hat{S}_{x})\hat{\sigma}_{z}-(\hat{S}_{x}\hat{S}_{y}+\hat{S}_{y}\hat{S}_{x})\hat{\sigma}_{x}\right]
	\end{split}
	\end{equation}
	which is equivalent of Eq.~(4) in the main text, using the transformations $\hat{S}_\pm=\hat{S}_x\pm i\hat{S}_y$ and $\hat{\sigma}_{\pm}=(\hat{\sigma}_{x}\pm i\hat{\sigma}_{y})/2$.
	By comparing the structures of Eqs.~\eqref{eq:SM:1} and \eqref{eq:SM:2}, and evaluating $\langle e_{x}|\hat{H}_\text{orb}|e_{x}\rangle$ and $\langle e_{y}|\hat{H}_\text{orb}|e_{y}\rangle$, we arrive at
	\begin{equation}
	\begin{split}
	\label{eq:SM:3}
	D_{1}=D_{xz}^{(xx)}=D_{yz}^{(yy)} 
	\qquad \text{and} \qquad \qquad \\ 
	D_{2}=\frac{D_{yy}^{(xx)}-D_{xx}^{(xx)}}{2}=\frac{D_{xx}^{(yy)}-D_{yy}^{(yy)}}{2} \text{,}
	\end{split}
	\end{equation}
	showing a practical method for directly evaluating $D_1$ and $D_2$ from Kohn-Sham DFT calculations. 
	We note that an arbitrary $\varphi$ in $\langle e_{\varphi}|\hat{H}_\text{orb}|e_{\varphi}\rangle$ leads to the following expressions: 
	\begin{equation}
	\begin{split}
	\label{eq:SM:4}
	|D_{1}|=\sqrt{(D_{xz}^{(\varphi)})^{2}+(D_{yz}^{(\varphi)})^{2}}
	\qquad\text{and} \qquad \\
	|D_{2}|=\sqrt{(D_{xy}^{(\varphi)})^{2}+\frac{1}{4}(D_{xx}^{(\varphi)}-D_{yy}^{(\varphi)})^{2}}\text{.} \qquad
	\end{split}
	\end{equation}
	
	For instance, DFT calculation at the PBE level, performed with $(a_{1\downarrow}^{1}a_{1\uparrow}^{0}e_{x\downarrow}^{1}e_{y\downarrow}^{1}e_{x\uparrow}^{0.5}e_{y\uparrow}^{0.5})$ occupation constraints (see the next Appendix~\ref{app:F} for details), results in the following zero-field splitting tensors:
	\begin{equation}
	\begin{array}{c}
	\label{eq:SM:5}
	\qquad \boldsymbol{D}_{\text{cartesian}}^{(xx)}=\begin{pmatrix}237 & -254 & 932\\
	-254 & 239 & 927\\
	931 & 927 & -476
	\end{pmatrix}\text{MHz}
	\\ \text{and} \qquad
	\boldsymbol{D}_{\text{cartesian}}^{(yy)}=\begin{pmatrix}-237 & 1323 & 138\\
	138 & -239 & 142\\
	138 & 142 & 475
	\end{pmatrix}\text{MHz}
	\end{array}
	\end{equation}
	expressed in the Cartesian coordinate system of the diamond lattice.
	Transforming into the local coordinate system defined by $\boldsymbol{v}_x^{T}=( \overline{1} \, \overline{1} \, 2 )/\sqrt{6})$, $\boldsymbol{v}_y^{T}=( 1  \, \overline{1} \, 0 )/\sqrt{2}$, $\boldsymbol{v}_z^{T}=(1 \, 1 \, 1)/\sqrt{3}$, using the rotation matrix $\boldsymbol{R}= [ \boldsymbol{v}_x$ $\boldsymbol{v}_y$ $\boldsymbol{v}_z ]$ of the diamond NV center, yields the following matrices (where numerical inaccuracies leading to small values such as $+1$ and $\pm3$ are ignored):
	\begin{widetext}
		\begin{equation}
		\label{eq:SM:6}
		\begin{split}\boldsymbol{D}_{\text{DFT}}^{(xx)}=\boldsymbol{R}^{-1}\boldsymbol{D}_{\text{cartesian}}^{(xx)}\boldsymbol{R}=\begin{pmatrix}-\frac{1}{3}D-D_{2} & 0 & +D_{1}\\
		0 & -\frac{1}{3}D+D_{2} & 0\\
		+D_{1} & 0 & +\frac{2}{3}D
		\end{pmatrix}=\begin{pmatrix}-1561 & +3 & +221\\
		+3 & +492 & +1\\
		+221 & +1 & +1069
		\end{pmatrix}\text{MHz}\text{,}\\
		\\
		\boldsymbol{D}_{\text{DFT}}^{(yy)}=\boldsymbol{R}^{-1}\boldsymbol{D}_{\text{cartesian}}^{(yy)}\boldsymbol{R}=\begin{pmatrix}-\frac{1}{3}D-D_{2} & 0 & +D_{1}\\
		0 & -\frac{1}{3}D+D_{2} & 0\\
		+D_{1} & 0 & +\frac{2}{3}D
		\end{pmatrix}=\begin{pmatrix}+492 & -3 & -221\\
		-3 & -1561 & -1\\
		-221 & -1 & +1069
		\end{pmatrix}\text{MHz}\text{,}
		\end{split}
		\end{equation}
	\end{widetext}
	From these results, the \textit{ab initio} zero-field splitting parameters are determined as:
	\begin{equation}
	\label{eqapp:Dpars}
	\begin{split}
	D_{\text{DFT}}=D_{zz}^{(xx)}-(D_{xx}^{(xx)}+D_{yy}^{(xx)})/2=+1603\:\text{MHz} \text{,}
	\\
	D_{1,\text{DFT}}=D_{xz}^{(xx)}=+221\:\text{MHz} \text{,}
	\\
	D_{2,\text{DFT}}=(D_{yy}^{(xx)}-D_{xx}^{(xx)})/2=+1027\:\text{MHz} \text{.}
	\end{split}
	\end{equation}
	However, the values of \(D_1\) and \(D_2\) are further reduced by the Ham reduction factor \(q=0.631\), as discussed in the main text:
	\begin{equation}
	\begin{split}
	D_{\text{theory}}=D_{\text{DFT}}=+1603\:\text{MHz} \text{,}
	\\
	D_{1,\text{theory}}=q \times D_{1,\text{DFT}}=+140\:\text{MHz} \text{,}
	\\
	D_{2,\text{theory}}=q \times D_{2,\text{DFT}}=+648\:\text{MHz} \text{.}
	\end{split}
	\end{equation}
	These final theoretical values are now in agreement with experimental data. We note that the spin-orbit parameter in Eq.~\eqref{eq:FullHamilton} is influenced by \(p=0.264\) thus the final value will be $\lambda_{\text{theory}}= p \lambda_{\text{DFT}} = 0.257 \times 20.4$~GHz $ = 4.1$~GHz in HSE06 calculations as discussed in Ref.~\onlinecite{Thiering2017}. We note that we recalculated the Jahn-Teller parameters where $\hbar\omega = 77.6$~meV of the harmonic oscillator was derived from a Huang-Rhys spectrum in Ref.~\onlinecite{Thiering2017} rather than taken from a fit to the paraboloid of the HSE06 APES which results in $\hbar\omega = 66.3$~meV. The latter mode is the correct procedure which modifies the original $p=0.304$ in Ref.~\onlinecite{Thiering2017} to $p=0.264$. We provide a Python script and input-output files to conveniently evaluate Jahn-Teller parameters in the supporting data.
	
	One may wonder that within a DFT calculation with $C_{3v}$ symmetry constraint and enforced $(\dots{e_{x\uparrow}^{0.5}}{e_{y\uparrow}^{0.5}})$ occupation $|e_x\rangle$ and $|e_y\rangle$ orbitals may freely mix with each other. Thus, any arbitrary linear combination of these orbitals ($|e_{\varphi}\rangle=\cos(\frac{\varphi}{2})|e_{x}\rangle-\sin(\frac{\varphi}{2})|e_{y}\rangle$) can be the convergent result of such DFT electronic cycle. When this happens evaluation of $D_{1,2}$, $A_{1,2}$ , $P_{1,2}$ parameters by means of Eq.~\eqref{eqapp:Dpars} is not possible. First, it seems convenient to use the "Pythagorean theorem-like expressions that of Eq. \eqref{eq:SM:4}. However, the square root therein renders the sign of all $D_{1,2}$, $A_{1,2}$ , $P_{1,2}$ parameters ambiguous. Therefore, one may forcefully symmetrize the orbitals by means of the generalized projection operators that of Eq.~\eqref{eq:Projection}. However, we use a more straightforward method by utilizing one of the three Jahn-Teller distorted minima exhibiting $C_{1h}$ symmetry where $|e_x\rangle$, $|e_y\rangle$ orbitals are stabilized by symmetry distortion. We start a DFT calculation on a relaxed structure with $C_{3v}$ symmetry constraint imposed on atomic positions but use the previously converged Jahn-Teller distorted electronic wavefunctions as a starting point. We enforce $(\dots{e_{x\uparrow}^{0.5}}{e_{y\uparrow}^{0.5}})$ occupation during the electronic self consistent cycle (SCF) and thus we obtain pure $|e_x\rangle$ and $|e_y\rangle$ orbitals numerically. We note that $(\dots{e_{x\uparrow}^{0.5}}{e_{y\uparrow}^{0.5}})$ partial occupation during this step is necessary to this end. We observe that $(\dots{e_{y\uparrow}^{1}}{e_{x\uparrow}^{0}})$ occupation together with $C_{3v}$ constraints on the atomic positions leads to spontaneous mixing of $|e_x\rangle$ and $|a_1\rangle$ orbitals, see the last row in Table~\ref{tab:functest}.
	
	\section{DFT results with different functionals}
	\label{app:F}
	We compared our results using the PBE and HSE06 functionals. Based on our previous findings, the HSE06 functional generally provides better accuracy for electric field gradients and spin density, which are relevant for the $\boldsymbol{P}$ and $\boldsymbol{A}$ tensors, respectively.
	However, for the spin-spin $\boldsymbol{D}$ tensor, we opted for the PBE functional, as it yielded better agreement with experimental data compared to HSE06.
	Nevertheless, in this section, we provide the raw data for the spin Hamiltonians obtained using both functionals and different symmetry constraints, as summarized in Table~\ref{tab:functest}, which supplements Table~\ref{tab:parameters} of the main text. We note that all $D_1$, $D_2$, $A_1$, $A_2$, $P_1$, and $P_2$ parameters are scaled by the same reduction factor $q=0.652$, derived from HSE06 functional.
	
	The electronic structure was determined as follows: 
	first, we optimized the atomic positions of the excited state by converging the $(a_{1\downarrow}^{1}a_{1\uparrow}^{0}e_{x\downarrow}^{1}e_{y\downarrow}^{1}e_{x\uparrow}^{0.5}e_{y\uparrow}^{0.5})$ single-electron occupation, which corresponds to the $(a_{1\uparrow}^{1}e_{x\uparrow}^{0.5}e_{y\uparrow}^{0.5})$ two-hole (two-particle) system.
	As a result, we obtained a solution that respects $C_{3v}$ symmetry in both atomic positions and electronic orbitals.
	Finally, we enforced hole occupancy in either $(a_{1\uparrow}^{1}e_{x\uparrow}^{1})$ or $(a_{1\uparrow}^{1}e_{y\uparrow}^{1})$ configurations to determine the orbital-dependent spin tensors: ($\boldsymbol{D}_{\text{DFT}}^{(xx)}$, $\boldsymbol{A}_{\text{DFT}}^{(xx)}$, $\boldsymbol{P}_{\text{DFT}}^{(xx)}$) or ($\boldsymbol{D}_{\text{DFT}}^{(yy)}$ , $\boldsymbol{A}_{\text{DFT}}^{(yy)}$, $\boldsymbol{P}_{\text{DFT}}^{(yy)}$), respectively. These computations were carried out using the \textsc{vasp} code, with the corresponding results provided in the first two rows of Table~\ref{tab:functest} labeled as \textsuperscript{\ref{Sup1}}.
	
	However, we also allowed the $(a_{1\uparrow}^{1}e_{x\uparrow}^{1})$ system to relax into its JT minimum, which lowered the symmetry to $C_{1h}$, allowing mixing between $|a_1\rangle$ and $|e_x\rangle$ orbitals as they both belong to the $|a^{\prime}\rangle$ representation in the reduced $C_{1h}$ symmetry. Notably, $|e_x\rangle$ transforms as $|a^{\prime}\rangle$ since it does not change its sign under $\hat{\sigma}_v^{(1)}$ mirror symmetry operation, as illustrated in Fig.~\ref{fig:APES}(b) of the main text. We note that $|e_y\rangle$ would become $|a^{\prime\prime}\rangle$ as it does change its sign under $\hat{\sigma}_v^{(1)}$.
	We found that all $D_1$, $D_2$, $A_1$, $A_2$, $P_1$, and $P_2$ parameters increased significantly when symmetry constraints were lifted (see the respective rows of Table~\ref{tab:functest} labeled as \textsuperscript{\ref{Sup2}}).
	However, terms that are not affected by orbital degeneracy ($D$, $A_\parallel$, $A_\perp$, $P$) exhibited much smaller changes due to symmetry lowering.
	
	Interestingly, this effect persisted even when the atomic positions were constrained to $C_{3v}$ symmetry while allowing single-particle levels to relax to $C_{1h}$ symmetry. Further, fully removing symmetry constraints, the last row of Table~\ref{tab:functest} labeled as \textsuperscript{\ref{Sup3}}, resulted in a 1.1~meV lower total energy with the PBE functional, while maintaining the same atomic positions as those in the PBE $C_{1h}$ row (\textsuperscript{\ref{Sup2}}) of Table~\ref{tab:functest}.
	
	In summary, whether to include or exclude orbital contamination between the $|a_{1}\rangle$ and $|e_{x}\rangle$ levels remains an open question, potentially warranting further investigation. However, we found the best agreement with experimental data for the parameters $D^{(e)}, D_1^{(e)}, D_2^{(e)}$ when enforcing $C_{3v}$ symmetry on both atomic positions and orbitals, and using the PBE functional. Consequently, we employed this method to evaluate the unknown parameters $A_{1}^{(e)}$, $A_{2}^{(e)}$, $P^{(e)}$, $P_1^{(e)}$, and $P_2^{(e)}$.
	
	\begin{table*}
		\caption{\label{tab:functest} Comparison of the \textit{ab initio} spin tensors obtained using the HSE06 and PBE functionals. The values highlighted in bold are included in Table~\ref{tab:parameters} of the main text.}
		\begin{ruledtabular} 
			\begin{tabular}{cc|c|ccc|cccc|ccc}
				&  & $\lambda^{(e)}$ & $D^{(e)}$ & $D^{(e)}_{1}$ & $D^{(e)}_{2}$ & $A^{(e)}_{\parallel}$ & $A^{(e)}_{\perp}$ & $A^{(e)}_{1}$ & $A^{(e)}_{2}$ & $P^{(e)}$ & $P^{(e)}_{1}$ & $P^{(e)}_{2}$ \tabularnewline
				func.\footnote{\label{SupA}DFT functional that we use.}  &  sym.  & (GHz) & (MHz) & (MHz) &  (MHz) & (MHz) & (MHz) &  (kHz) & (kHz) & (MHz) & (kHz) & (kHz)\tabularnewline
				\hline 
				
				expt.\footnote{\label{Sup0}References for experimental data are depicted in Table~\ref{tab:parameters} of the main text.} &  & \textbf{5.3} & \textbf{1420} & \textbf{141}  & \textbf{775} & \textbf{-40} & \textbf{-23}  & n. a. & n. a. & n. a. & n. a. & n. a.\tabularnewline
				PBE & $C_{3v}$\footnote{\label{Sup1}Single particle orbitals and excited state atomic positions are converged with $(a_{1\downarrow}^{1}a_{1\uparrow}^{0}e_{x\downarrow}^{1}e_{y\downarrow}^{1}e_{x\uparrow}^{0.5}e_{y\uparrow}^{0.5})$ occupation to conserve orbital symmetry.} 
				& 5.33\footnote{\label{lambdaSupp}Extrapolated value towards infitely large supercell multiplied by $p=0.262$, see Fig. 1. in Supp. Mat. that of Ref.~\onlinecite{Thiering2017} for details. } & \textbf{1603} & \textbf{140}  & \textbf{648} & \textbf{38.6}  & \textbf{24.8}  & \textbf{-55.8} & \textbf{-43.5} & \textbf{-3.8}  & \textbf{10.4}  & \textbf{8.2} \tabularnewline
				HSE06 & $C_{3v}$\textsuperscript{\ref{Sup1}} & 
				4.13\textsuperscript{\ref{lambdaSupp}} & 2321 & 140 & 733 & 40.1  & 26.6  & -57.8 & -45.3 & -3.9  & 10.8  & 8.5 \tabularnewline
				PBE & $C_{1h}$\footnote{\label{Sup2}Single particle orbitals and excited state atomic positions are converged without any symmetry constraint. The result follows $C_{1h}$ symmetry with $(a_{\downarrow}^{\prime1}a_{\uparrow}^{\prime0}a_{\downarrow}^{\prime1}a_{\downarrow}^{\prime\prime1}a_{\uparrow}^{\prime0}a_{\uparrow}^{\prime\prime1})$ occupation. } & $-$& 1665 & 280 & 872 & 40.1  & 25.0  & -125.2 & -45.8 & -3.8  & 9.0  & 14.2 \tabularnewline
				HSE06 & $C_{1h}$\textsuperscript{\ref{Sup2}} 
				& $-$ & 2278 & 343 & 1063 & 40.8  & 27.0  & -89.2 & -47.3 & -3.9  & 16.8  & 16.6 \tabularnewline
				PBE & $C_{1h}$\footnote{\label{Sup3}The atomic positions are taken from the PBE row labeled as \textsuperscript{\ref{Sup1}}, ensuring that the system's geometry adheres to $C_{3v}$ symmetry. However, the single-particle levels are optimized without any symmetry constraints, resulting in a $C_{1h}$ symmetry, as observed in the rows labeled \textsuperscript{\ref{Sup2}}.  } 
				&$-$& 1657  & 286  & 881 & 40.3  & 26.3 & -125.6 & -46.2 & -3.7  & 9.3  & 14.6 \tabularnewline
			\end{tabular}
		\end{ruledtabular} 	
	\end{table*}

	\section{Coherence time during optical readout}
	\label{app:G}
	We assume that nuclear spin coherence is limited by thermal excitations between $|E_x\rangle$ and $|E_y\rangle$ induced by acoustic phonons. These states belong to the $m_S=0$ manifold and are split by $\delta_\perp \approx 3$~GHz due to strain. Following Eq.~(5) of Ref.~\onlinecite{Jahnke_2015}, the phonon-induced orbital relaxation rates are:
	\begin{equation}
	\label{eq:Coh1}
	\gamma_{+}=C\times\delta_{\perp}^{3}n\quad\text{ and}\quad\gamma_{-}=C\times\delta_{\perp}^{3}(n+1)\text{,}
	\end{equation}
	where 
	\begin{equation}
	\label{eq:Coh2}
	n=\frac{1}{\exp\bigl(\frac{\delta_{\perp}h}{k_{B}T}\bigr)-1}.
	\end{equation}
	The term $n$ represents the thermal population of acoustic phonons at frequency $\delta_\perp$. For simplicity, we use parameters from the neutral NV defect with similar orbitals with those of the negatively charged NV defect, which exhibits a relaxation time of $\tau_-^{-1} = 0.43$~$\mu$s at 4.6 K, with its $|^2E\rangle$ level split by $\delta_\perp = 9.8$~GHz.
	By scaling Eq.~\eqref{eq:Coh1} to the $|E_{x,y}\rangle$ excited states of NV center at 5~K used during the experiments~\cite{Monge_2023}, we estimate a coherence limit of $\tau_+ \approx 4.5$~$\mu$s for manipulation using the $P_2^{(e)}$ term in Eq.~\eqref{eq:HssE}. Notably, this value is of the same order of magnitude as the coherence times ($T^*_2  \approx 2 ... 7$~$\mu$s) assumed in Section 2 of the Supplemental Materials in Ref.~\onlinecite{Monge_2023}. Consequently, we adopt a heuristic approximation of $T^*_2=4 \pm 2$~$\mu$s for coherence time in the preceding section.
	
	We note that hyperfine transitions involving $A_\perp^{(e)}$ lose coherence significantly faster. This rapid decoherence occurs during every consecutive optical cycle due to the different temporal phase shifts ($e^{-i2\pi D^{(g)}t}$) for $m_S=\pm1$ and $m_S=0$ in the $|^3 A_2\rangle$ ground state. The characteristic coherence time in this case is approximately: ${T_{2}^{*}} \approx (D^{(g)})^{-1}=(2.88~$GHz$)^{-1} = 0.3 $~ns.
	This 0.3~ns timescale is much shorter than both the radiative lifetime ($\gamma_\text{rad} =12$~ns) and the waiting time ($\gamma_\text{wait}=0.5$~ns), which supports the $2 \times n$ repeated incoherent processes described in Eq.~\eqref{eq:hyperfine}.
	Additionally, we emphasize that this type of decoherence does not occur in $P_2^{(e)}$ driving, where the system remains entirely within the $m_S = 0$ spin projection.
	
	\section{Interpretation of the Fermi golden rule during optical cycles}
	\label{app:H}
	In this section we will describe nuclear spin flip process introduced by $\frac{1} {2}A^{(e)}_{\perp}\big(\hat{S}_{+}\hat{I}_{-}+\hat{S}_{-}\hat{I}_{+}\big)$ hyperfine interaction. For simplicity, we will disregard the orbital degrees of freedom ($m_L$) in our discussion. Therefore, we initialize the system in the $|^3 A_2^{m_S=0}\rangle$ ground state multiplet with nuclear spin $m_I=\pm1$ that we label as
	\begin{equation}
	\label{eq:Psignd}
	\Psi_{\text{gnd}}=\overset{m_{S}}{\overbrace{|0\rangle}}\otimes\overset{m_{I}}{\overbrace{|\pm1\rangle}} \text{,}
	\end{equation}
	that we will excite into the $|^3 E \rangle$ manifold that of Eq.~\eqref{eq:3EWave}. The absorption of the photon cannot flip the electronic spin ($m_S$) or the nuclear spin ($m_I$) either thus it should stay in $|0\rangle\otimes|\pm1\rangle$ even within $|^3 E \rangle$. Meanwhile, $A^{(e)}_{\perp}$ hyperfine can be interpreted as an external perturbation as long as the system resides in the upper $|^3 E\rangle$ multiplet. Thus, we introduce $\tau_\text{Rabi}=(D^{(e)})^{-1}=(1.42$~GHz$)^{-1}=$~0.7~ns as the Rabi oscillation frequency between $|0\rangle\otimes|\pm1\rangle$ and
	$|\pm1\rangle\otimes|0\rangle$ states induced by $A^{(e)}_{\perp}$ where $D^{(e)}$ is the energy spacing between $m_S=\pm1$ and $m_S=0$ spin projections (zero-field splitting). We graphically depict this oscillation ($\tau_\text{Rabi}=$~0.43~ns) in Fig.~\ref{fig:hyper}(a) where we additionally take into account the orbitals ($m_L$) and strain splitting that of Eq.~\eqref{eq:hyperfine}. At this point, one can notice that the excited state lifetime ($\tau_{\text{rad}}$) is much longer than this Rabi frequency: $\tau_{\text{Rabi}}\ll\tau_{\text{rad}}\approx 12$~ns thus Fermi golden rule remains valid for entanglement induced by $A_{\perp}^{(e)}$.
	
	Therefore, we diagonalize the hyperfine interaction related problem by
	\begin{equation}
	\tag{H2}
	\begin{split}
	\label{eq:Psiex}
	\Psi_{\text{ex}}^{0}=\sqrt{1-\delta^{2}}|0\rangle\otimes|\pm1\rangle+\delta|\pm1\rangle\otimes|0\rangle \text{,} \\
	\Psi_{\text{ex}}^{\pm}=\sqrt{1-\delta^{2}}|\pm1\rangle\otimes|0\rangle-\delta|0\rangle\otimes|\pm1\rangle \text{,}
	\end{split}
	\end{equation}
	where $\delta={A_{\perp}/D^{(e)}}$ is the small perturbation. There are four consecutive absorption followed by emission processes possible
	\begin{equation}
	\tag{H3}
	\label{eq:Golden}
	\begin{aligned}\boxed{\text{I}}\;p\left(\Psi_{\text{gnd}}\!\rightarrow\!\Psi_{\text{ex}}^{0}\!\rightarrow\!\Psi_{\text{gnd}}\right) & \!=\!\!\!\! & (1-\delta^{2})^{2}\approx & \,1-2\delta^{2}\!\!\!\!\!\!\!\!\!\!\!\!\\
	\boxed{\text{II}}\;p\left(\Psi_{\text{gnd}}\!\rightarrow\!\Psi_{\text{ex}}^{0}\!\rightarrow\!\Psi_{\text{gnd}}^{*}\right) & \!=\!\!\!\! & (1-\delta^{2})\delta^{2}\approx & \,\delta^{2}\\
	\text{\boxed{\text{III}}\;}p\left(\Psi_{\text{gnd}}\!\rightarrow\!\Psi_{\text{ex}}^{\pm}\!\rightarrow\!\Psi_{\text{gnd}}\right) & \!=\!\!\!\! & \delta^{2}\delta^{2}\quad\approx & \,0\\
	\text{\boxed{\text{IV}}}\;p\left(\Psi_{\text{gnd}}\!\rightarrow\!\Psi_{\text{ex}}^{\pm}\!\rightarrow\!\Psi_{\text{gnd}}^{*}\right) & \!=\!\!\!\! & \delta^{2}(1-\delta^{2})\approx & \,\delta^{2} \text{,}
	\end{aligned}
	\end{equation}
	where $\Psi_{\text{gnd}}^{*}=|\pm\rangle\otimes|0\rangle$ is final ground state where the nuclear spin is relaxed to $m_I=0$. We note that the two processes treated with Fermi golden rule are independent because there is no correlation between the absorbed and emitted photons. This is in stark contrast with the nuclear spin flip process induced by $P_{2}^{(e)}\bigl[\hat{I}_{-}^{2}\hat{\sigma}_{-}\!+\hat{I}_{+}^{2}\hat{\sigma}_{+}\bigr]$ in Fig.~4(b) where $\tau_{\text{rad}}\approx 12$ is overshadowed by $\tau_\text{Rabi}=$~55~$\mu$s justifying the coherent driving by $P_{2}^{(e)}$.
	
	One may observe that the nuclear spin flip fundamentally occurs in emission for process II while it occurs in absorption for process III. Thus, Eq.~\eqref{eq:perturbation_A_perp} incorporates only the first absorption step justifying the 2$\times$ factor in Eq.~\eqref{eq:hyperfine}.

	\newpage

	%
	
\end{document}